\documentclass[3p]{elsarticle}
\usepackage{amsmath}
\usepackage{graphicx}

\usepackage{enumitem}

\usepackage{multirow}
\usepackage{amsmath}
\usepackage{cancel}
\usepackage{amssymb}
\usepackage{amsfonts}
\usepackage{array}
\usepackage{tcolorbox}
\usepackage{adjustbox}

\usepackage{subfig}
\usepackage{algorithmicx,float}
\usepackage{subfig}
\usepackage[textsize=tiny]{todonotes}

\newtheorem{theorem}{Theorem}

\usepackage{xr}

\usepackage{tcolorbox}
\usepackage[linesnumbered,ruled]{algorithm2e}
\newenvironment{proc}[1][htb]
  {
   \begin{algorithm}%
  }{\end{algorithm}}
\usepackage{hyperref}
\usepackage{mathtools}
\usepackage{float}
\usepackage{lineno,hyperref}
\modulolinenumbers[5]

\journal{Journal of \LaTeX\ Templates}

\begin{document}

\begin{frontmatter}

\title{ Griefing-Penalty: Countermeasure for Griefing Attack in Lightning Network }
\tnotetext[mytitlenote]{A preliminary version of our paper was accepted in the proceedings of The 19th IEEE International Conference on Trust, Security and Privacy in Computing and Communications (IEEE TrustCom 2020) under the title "Time is Money: Time is Money: Countering Griefing Attack in Lightning Network" \cite{agrief}.}


\author[mymainaddress]{Subhra~Mazumdar\corref{mycorrespondingauthor}}
\cortext[mycorrespondingauthor]{Corresponding author}
\ead{subhra.mazumdar1993@gmail.com}

\author[mymainaddress,mythirdaddress]{Prabal~Banerjee}
\ead{mail.prabal@gmail.com}

\author[mymainaddress,mysecondaryaddress]{Sushmita~Ruj}
\ead{Sushmita.Ruj@data61.csiro.au}

\address[mymainaddress]{Cryptology and Security Research Unit, Indian Statistical Institute, Kolkata, India}
\address[mysecondaryaddress]{ CSIRO Data61, Australia}
\address[mythirdaddress]{Polygon (previously Matic Network)}
\begin{abstract}
Lightning Network can execute unlimited number of off-chain payments, without incurring the cost of recording each of them in the blockchain. However, conditional payments in such networks is susceptible to \emph{Griefing Attack}. In this attack, an adversary doesn't resolve the payment with the intention of blocking channel capacity of the network. We propose an efficient countermeasure for the attack, known as \emph{Griefing-Penalty}. If any party in the network mounts a griefing attack, it needs to pay a penalty proportional to the \emph{collateral cost} of executing a payment. The penalty is used for compensating affected parties in the network. We propose a new payment protocol \emph{HTLC-GP} or \emph{Hashed Timelock Contract with Griefing-Penalty} to demonstrate the utility of the countermeasure. Upon comparing our protocol with existing payment protocol \emph{Hashed Timelock Contract}, we observe that the average revenue earned by the attacker decreases substantially for \emph{HTLC-GP} as compared to \emph{HTLC}. We also study the impact of path length for routing a transaction and rate of griefing-penalty on the budget invested by an adversary for mounting the attack. The budget needed for mounting griefing attack in \emph{HTLC-GP} is 12 times more than the budget needed by attacker in \emph{HTLC}, given that each payment instance being routed via path length of hop count 20. 
\end{abstract}


\end{frontmatter}

\section{Introduction}
Since the inception of Bitcoin \cite{bitcoin} in 2009, Blockchain technology has seen widespread adoption in the payment space. In spite of many desired features like decentralization and pseudonymity, a constant criticism faced by Bitcoin and Ethereum \cite{ethereum} is that of scalability. 

Layer-two protocols \cite{gudgeon2019sok} enables users to perform transactions \emph{off-chain}, massively cutting down data processing on the blockchain. \textit{Payment Channel} \cite{decker2015fast}, \cite{poon2016bitcoin} stood out as a practically deployable solution. It is modular in nature, without requiring any fundamental changes in the protocol layer. 
Except for the opening and closing of the payment channel, several transactions can be executed off-chain without recording it in Blockchain. In case of dispute, any party can unilaterally broadcast the latest valid transaction in the blockchain and terminate the channel. A malicious party will lose all the funds locked in the channel if it tries to broadcast any older transaction. Since opening a payment channel has its overhead in terms of time and funds locked, parties that are not connected directly leverage on the set of existing payment channels for transfer of funds. The set of payment channels form the \textit{Payment Channel Network} or PCN \cite{poon2016bitcoin}. In practice, Lightning
Network for Bitcoin \cite{poon2016bitcoin} and Raiden Network for Ethereum \cite{raiden} are the widely deployed PCNs.

  \begin{figure}[!ht]
    \centering
    \includegraphics[width=9cm]{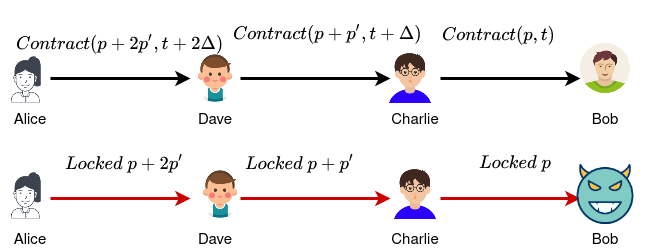}
    \caption{\emph{Bob} mounts Griefing Attack }
    \label{bob}
\end{figure}

\paragraph*{\textbf{Payment in Lightning Network}} If a sender and a recipient of payment do not share a channel, such payments make use of Hashed Timelock Contracts or HTLCs \cite{poon2016bitcoin}. Since the payment gets routed via several intermediaries, a specific condition is imposed via off-chain contracts in order to prevent cheating. Payments are contingent to fulfillment of this condition. We describe with an example how conditional payments get executed between parties not directly connected by payment channel in Lightning Network. Suppose \emph{Alice} wants to transfer $p$ coins to \emph{Bob} via path comprising payment channels \emph{Alice-Dave, Dave-Charlie} and \emph{Charlie-Bob}, as shown in Fig. \ref{bob}. Each intermediate node charge a processing fee of $p'$. \emph{Alice} forwards a conditional payment to \emph{Dave}, forming an off-chain contract, denoted as \emph{Contract$(p+2p',t+2\Delta)$}, locking $p+2p'$ coins for a time period $t+2\Delta$. Here $\Delta$ is the worst-case confirmation time for settling a transaction on-chain. \emph{Dave} deducts $p'$ coins from the amount and forwards the payment to \emph{Charlie} by forming a off-chain contract, locking $p+p'$ coins for $t+\Delta$. Finally, \emph{Charlie} deducts $p'$ coins from the payment amount and locks $p$ coins with \emph{Bob} for a time period $t$. In order to claim $p$ coins from \emph{Charlie}, \emph{Bob} must resolve the payment within time period $t$. If the time period elapses, \emph{Charlie} goes on-chain to claim refund, closes the channel \emph{Charlie-Bob} and unlocks the money from the contract. Using the information released by \emph{Bob}, rest of the intermediaries resolve the payment as well, each claiming a processing fee of $p'$. 

\paragraph*{\textbf{Payment susceptible to Griefing Attack}} Griefing Attack in Lightning Network was first mentioned in \cite{robinson2019htlcs}. Paralyzing the network for multiple days by overloading each channel with maximum unresolved HTLCs has been studied in \cite{mizrahi2020congestion}, \cite{tochner2019hijacking}. In \cite{bank}, sybil nodes initiate several payments via multiple paths and griefs them simultaneously. 

In the example described above, \emph{Bob} mounts griefing attack by not responding, as shown in Fig. \ref{bob}. \emph{Charlie} can go on-chain and withdraw the coins locked in the contract only after the elapse of the contract's timeperiod. Thus \emph{Bob} manages to lock $\mathcal{O}(p)$ coins in each of the preceding payment channels for a timeperiod of $t$ units, without investing any money. Note that $t$ could be of the order of \emph{24 hours}. Hence for an entire day, none of the parties can utilize the amount locked in their respective off-chain contracts.

\paragraph*{\textbf{Motive behind Griefing Attack}} By mounting griefing attack, an adversary may try to achieve either of the objectives:
\begin{itemize}[leftmargin=*]
  \item Stalling network using self payment: The adversary controls the sender and receiver of several payment requests, blocking multiple intermediaries from accepting any other payments to be routed through it \cite{bank}, \cite{zmn}. In order to decrease the network throughput, an adversary may setup several \emph{Sybil nodes} at strategic positions across the PCN and amplify the damage by submitting several payment requests.
%
%
  
%

  \item Eliminating a competitor from the network: The adversary tries to eliminate a competitor and block all its existing channel's outgoing capacity \cite{egger2019atomic}, \cite{zmn}.The adversary sets the victim as an intermediate node in the path carrying out the self-payment.  The transaction value of self payment is equivalent to the victim's outgoing channel capacity, jamming all the channels of the victim node. The victim cannot utilize the fund until the adversary decides to claim the payment. As a consequence, several future payment request which could have been routed through the victim node now gets routed through the adversary. It reaps indirect economic benefit by claiming the processing fee for routing such transactions. 

In Fig. \ref{eliminate}, Node $B$ has outgoing channel with $A$ and $C$, each of capacity \emph{0.1 BTC} (each party having a balance of \emph{0.05 BTC}). Node $D$ has channel with \emph{A} and \emph{C}, each of capacity \emph{0.2 BTC}. It conducts self-payment of \emph{0.05 BTC}, in each direction. Upon griefing for \emph{24 hrs}, $D$ denies $B$ from accepting any transaction request. $A$ and $C$, having residual outgoing capacity of \emph{0.1 BTC} each in channel \emph{AD} and \emph{CD}, is now forced to route all the payments via D. 
\begin{figure}[!ht]
    \centering
    \includegraphics[width=6cm]{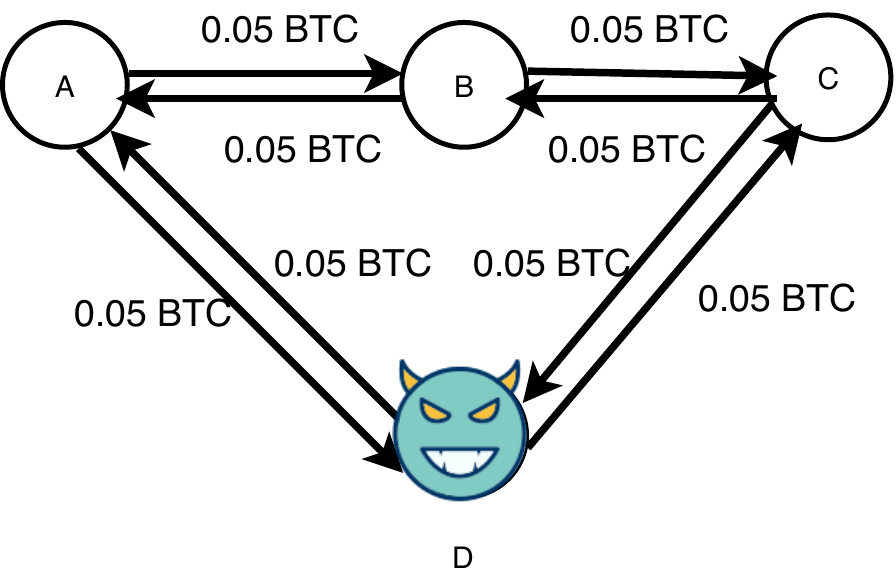}
    \caption{Eliminating a competitor}
    \label{eliminate}
\end{figure}

  \item Stalling network using intermediary: The adversary controls a node with a high degree centrality and broadcasts its processing fee to be extremely low in order to ensure multiple payments get routed through such nodes \cite{rohrer2019discharged}. It later ignores all the payment by not forwarding the message to outgoing neighbours, locking funds across multiple paths thereby affecting a large portion of the network.


\end{itemize}
\subsection{Our Goal}
\begin{figure}[!ht]
    \centering
    \includegraphics[width=9cm]{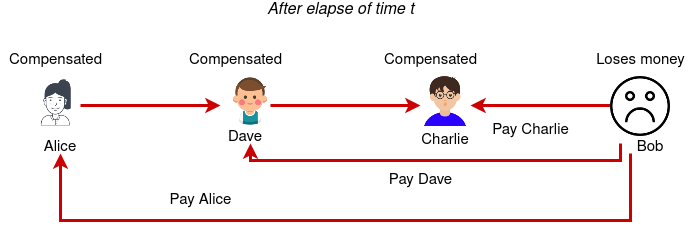}
    \caption{\emph{Bob} is penalized }
    \label{bob1}
\end{figure}

Griefing attack in Lightning Network cannot be prevented as long as a malicious node has nothing to lose or, in other words, it has \emph{nothing at stake}. The problem cannot be solved until and unless the attacker has the fear of losing money upon mounting the attack. Thus, before accepting the payment parties must make a commitment to pay a compensation, in case they stop responding intentionally. The amount deducted from adversary's balance must be able to compensate all the parties which got affected by the attack. A high level idea of the countermeasure has been pictorially depicted in Fig. \ref{bob1}. \emph{Alice} forwards the payment to \emph{Bob} via some intermediaries. Each party accepting the off-chain contract is supposed to lock an amount, which gets deducted if the party fails to resolve the payment before the contract timeout period. Here \emph{Bob} doesn't respond intentionally, allowing the timeout period of the contract to elapse.  As per the terms of the contract, he gets penalized and the funds slashed from his account is used to compensate \emph{Alice, Dave} and \emph{Charlie}.

\subsection{Our Contributions}
In this paper, we have made the following contributions:
\begin{itemize}[leftmargin=*]
\item We propose a countermeasure for mitigating griefing attack in Lightning Network, known as \emph{Griefing-Penalty}. It punishes the griefer by forcing it to pay compensation to all the parties whose funds got locked for a certain time-period as a result of the attack.
\item To illustrate the benefit of the proposed countermeasure, we propose a new payment protocol, called \emph{HTLC-GP} or Hashed Timelock Contract with Griefing-Penalty. The penalty deducted is a fraction of the amount of funds locked by the attacker per unit time. This fraction is termed as \emph{rate of griefing penalty}.
\item We propose a construction for secure multihop payment using HTLC-GP and provide security analysis for the same. It proves that our protocol is privacy preserving and mitigates loss due to griefing attack by compensating the honest nodes.


%
%
%
\item We study two attacking strategies for eliminating competitor node from network. Upon mounting the griefing attack following either of the strategy, we compare the profit made by the attacker in \emph{HTLC} and \emph{HTLC-GP}, by executing the protocols on several snapshots of Lightning Network. The profit here is termed as \emph{Return on Investment (RoI)}. It is observed that RoI is negative for \emph{HTLC-GP} compared to a positive RoI in \emph{HTLC}, hence disincentivizing a node from mounting griefing-attack due to substantial loss incurred. 
\item We compare the investment made by the adversary for mounting the attack in both \emph{HTLC-GP} and \emph{HTLC} upon varying \emph{path length} and \emph{rate of griefing-penalty}. The budget needed for mounting griefing attack in \emph{HTLC-GP} is 4 times more than the budget needed for HTLC, when the path length is set to 4. The ratio increases to 12 for path length of 20, which is the maximum hop count allowed in Lightning Network. For a fixed path length, the ratio increases upto 500 for rate of griefing penalty exceeding $10^{-3}$.
\end{itemize}


\subsection{Organization of the Paper}
 We provide provide a high level overview of our proposed countermeasure, \emph{Griefing-Penalty}, in Section \ref{strategy}. Based on this idea, we have proposed a new payment protocol, \emph{HTLC-GP} or \emph{Hashed Timelock Contract with Griefing-Penalty} in Section \ref{rg1}. We provide a detailed construction of Multi hop payment using HTLC-GP in Section \ref{paygp}. Security analysis of the proposed multihop payment protocol has been provided in Section \ref{secgp1}. We divide the \emph{Performance Evaluation} in Section\ref{performance} into two parts. Firstly, we analyze the profit earned by eliminating competitor in Section \ref{profit}, demonstrating the efficiency of \emph{HTLC-GP} over \emph{HTLC} in countering griefing attack. Next, we discuss in Section \ref{vary1} the impact of certain parameters on the investment made by attacker in HTLC-GP. Related Works has been stated in Section \ref{related} and finally, we conclude the paper stating the scope for future work in Section \ref{conclusion}. Notations used in the paper is given in Table \ref{tab:not}.

 \begin{table*}[h]
\centering
  

  \begin{tabular}{|c |c |} 
    \hline
 Notation &Description \\
 \hline
 
$G(V,E)$ &Graph representing the Lightning Network\\
$V$   &Set of nodes in Lightning Network\\
$E$   &Set of payment channels in Lightning Network, $E \subset V \times V$\\
$U_0$ & Payer/Sender, $U_0 \in V$\\
$U_n$ & Payee/Receiver, $U_n \in V$\\
  $\alpha$ & Amount to be transferred from $U_0$ to $U_n$\\
  $P$ &Path connecting $U_0$ to $U_n$\\
  $n$ & Length of the path $P$\\
  $U_i \in V, i \in [0,n]$ &Nodes in $P, (U_i,U_{i+1}) \in E$\\
  $locked(U_i,U_{j})$ &Amount of funds locked by $U_{i}$ in the payment channel $(U_i,U_{j})$\\
  $remain(U_i,U_j)$ &Net balance of $U_i$ that can be transferred to $U_j$ via off-chain transaction \\
 $fee(U_i)$ &Processing fee charged by $U_i$ for forwarding the payment\\
 $\lambda$ &Security Parameter\\
 $\mathcal{H}\{0,1\}^*\rightarrow \{0,1\}^\lambda$ & Standard Cryptographic Hash function\\
 $\Delta$ &Worst-case confirmation time when a transaction is settled on-chain\\
$\gamma$ &Rate of griefing penalty (per minute) \\

\hline

\end{tabular}
\vspace{0.2cm}
\caption{Notations used in the paper}
\label{tab:not}
\vspace{-0.2cm}
\end{table*}

\section{Background}

\subsection{\textbf{Payment Channel Network}}
\label{basic}
A Payment Channel Network or PCN is defined as a bidirected graph $G:=(V,E)$, where $V$ is the set of accounts dealing with cryptocurrency and $E$ is the set of payment channels opened between a pair of accounts. A PCN is defined with respect to a blockchain. Apart from the opening and closing of the payment channel, none of the transaction gets recorded on the blockchain. Upon closing the channel, cryptocurrency gets deposited into each user's wallet according to the most recent balance in the payment channel. Every node $ v\in V$ charges a processing fee $fee(v)$, for relaying funds across the network. Correctness of payment across each channel is enforced cryptographically by hash-based scripts \cite{poon2016bitcoin} or scriptless locking \cite{malavolta2019anonymous}. Each payment channel $(v_i,v_j)$ has an associated capacity $locked(v_i,v_j)$, denoting the amount locked by $v_i$ and $locked(v_j,v_i)$ denoting the amount locked by $v_j$. $remain(v_i,v_j)$ signifies the residual amount of coins $v_i$ can transfer to $v_j$. Suppose sender $S$, which is node $v_0$, wants to transfer amount $\alpha$ to $R$, which is node $v_n$ through a path $v_0\rightarrow v_1 \rightarrow v_2 \ldots \rightarrow v_n$, with each node $v_i$ charging a processing fee $fee(v_i)$. If $remain(v_i,v_{i+1})\geq \alpha_i : \alpha_i=\alpha - \Sigma_{k=i}^n fee(v_{k}), i \in [0,n-1]$, then funds can be relayed across the channel $(v_i,v_{i+1})$. The residual capacity is updated as follows : $remain(v_i,v_{i+1})=remain(v_i,v_{i+1})-\alpha_i$ and $remain(v_{i+1},v_{i})=remain(v_{i+1},v_{i})+\alpha_i$.

\emph{Lightning Network} (LN) \cite{poon2016bitcoin} is the most widely accepted Bitcoin-compatible PCN. Two parties willing to open a channel, lock funds in 2-of-2 multi-signature contract. A new commitment transaction is created by exchange of signatures if both the parties agree to update the state of the channel. Such transactions can be broadcasted anytime, if a party wants to unilaterally close a channel without requiring any further cooperation from the counterparty. To invalidate the previous transaction before creating a new one, a revocation mechanism stated in \cite{poon2016bitcoin} requires parties to share the secret keys used for signing such a transaction. When a party goes on-chain broadcasting his or her copy of transaction, a time window is imposed which prevents the spending of the funds immediately. The time window is enforced by using relative timelocks \cite{timelock}. The counterparty must react within this time window in order to punish the malicious party. The former uses the secret key of the revoked transaction to spend the entire fund locked in the channel within the given time-window. The malicious party loses its funds. 


\subsection{\textbf{Hashed Time-lock Contract}}
\label{htlcdef}
Multihop Payment in Lightning Network is enabled by the use of \textit{Hashed Time-lock Contract (\emph{HTLC})} \cite{poon2016bitcoin}. A payer $S$ wants to transfer funds to a payee $R$, using a network of channels across an $n$-hop route $(v_0,v_1,v_2,\ldots,v_n), S=v_0, R=v_n$. The payee $R$ creates a condition $y$ defined by $y=\mathcal{H}(\tilde{x})$ where $\tilde{x}$ is a random string and $\mathcal{H}$ is a random oracle \cite{bellare1993random}. The condition $y$ is shared with $S$. The condition is shared across the whole payment path. Between any pair of adjacent nodes $(v_i,v_{i+1})$, the hashed time-lock contract is defined by $HTLC(v_i,v_{i+1},y,b,t)$. It implies that $v_i$ locks $b$ units of fund in this contract. The amount locked can be claimed by party $v_{i+1}$ only if it releases the correct preimage $\tilde{x}: y=\mathcal{H}(\tilde{x})$ within timeout $t$. If $v_{i+1}$ doesn't release $\tilde{x}$ within time $t$, then $v_i$ settles the dispute on-chain by broadcasting the transaction. The channel between $v_{i}$ and $v_{i+1}$ is closed and $v_i$ unlocks the money from the contract. If $v_{i+1}$ releases the preimage $x$ then it can either broadcast the transaction on-chain or settle the contract off-chain. Upon off-chain settlement, the contract is invalidated by creating a new commitment transaction, with $b$ units being added to $v_{i+1}$'s account. 
\emph{HTLC} acts like a conditional payment which is forwarded by each of the intermediate parties until it reaches the payee. If the payee or any other intermediate node ignores to resolve incoming contract
request and waits for the expiration of the off-chain contract,
the funds remain locked in all the channels starting connecting payer to the attacker. Note that after the timeout
period, all the parties withdraw the fund locked in the contract.
The attacker manages to mount griefing attack without losing any money in the process.

\section{Key Idea of Griefing-Penalty}
\label{strategy}

Designing fair protocols on Bitcoin, where the adversary is forced to pay a mutually predefined monetary penalty to compensate for the loss of honest parties was first introduced by Bentov et al. \cite{claimOrRefund}. Inspired by this idea, we propose a countermeasure for griefing attack, \emph{Griefing-Penalty}, to solve the problem of griefing in the Lightning Network. The griefing-penalty imposed on an adversary for mounting griefing attack on a path of length, $n$ is proportional to the summation of collateral cost of each payment channel involved in routing. \emph{Collateral cost per payment channel} is defined as the product of the amount locked in the off-chain contract and the expiration time of the contract. The amount deducted per unit time from adversary's balance is fraction of the collateral locked. This fraction is termed as \emph{rate of griefing penalty} or $\gamma$.  The reason behind considering the expiration time of the contract for accounting griefing-penalty is to punish griefer for denying service to other participants in the path. 

In the next section, we discuss how to incorporate griefing-penalty into the existing payment protocol, Hashed Timelock Contract or \emph{HTLC}. Note that use of \textit{Griefing-Penalty} is independent of the cryptographic primitive used for the underlying payment protocol. It can be incorporated in \emph{Scriptless Scripts} as well \cite{malavolta2019anonymous}. 

\section{A Simple Protocol for countering Griefing Attack: \textit{HTLC1.0}}
\label{htlcsimple}

We incorporate \emph{Griefing-Penalty} into \emph{HTLC} \cite{poon2016bitcoin}. Let us rename the modified payment protocol as \textit{HTLC1.0}. It is assumed that for a given node, individual griefing-penalty earned upon elapse of locktime of the off-chain contract is less than the expected revenue earned by processing several transaction request within the given locktime. We assume that all the nodes in the network are rational, whose intention is to maximize the earning by remaining active in the network. Hence any such rational player would prefer to utilize their funds rather than earn penalty by reverse-griefing and keep their funds locked in a channel. Based on these assumptions, we define \emph{HTLC1.0} with an example.

\subsection{Construction of two-party off-chain Revocable \textit{HTLC1.0}}
\label{revoc}
\begin{figure*}[!ht]
    \centering
    \includegraphics[scale=0.7]{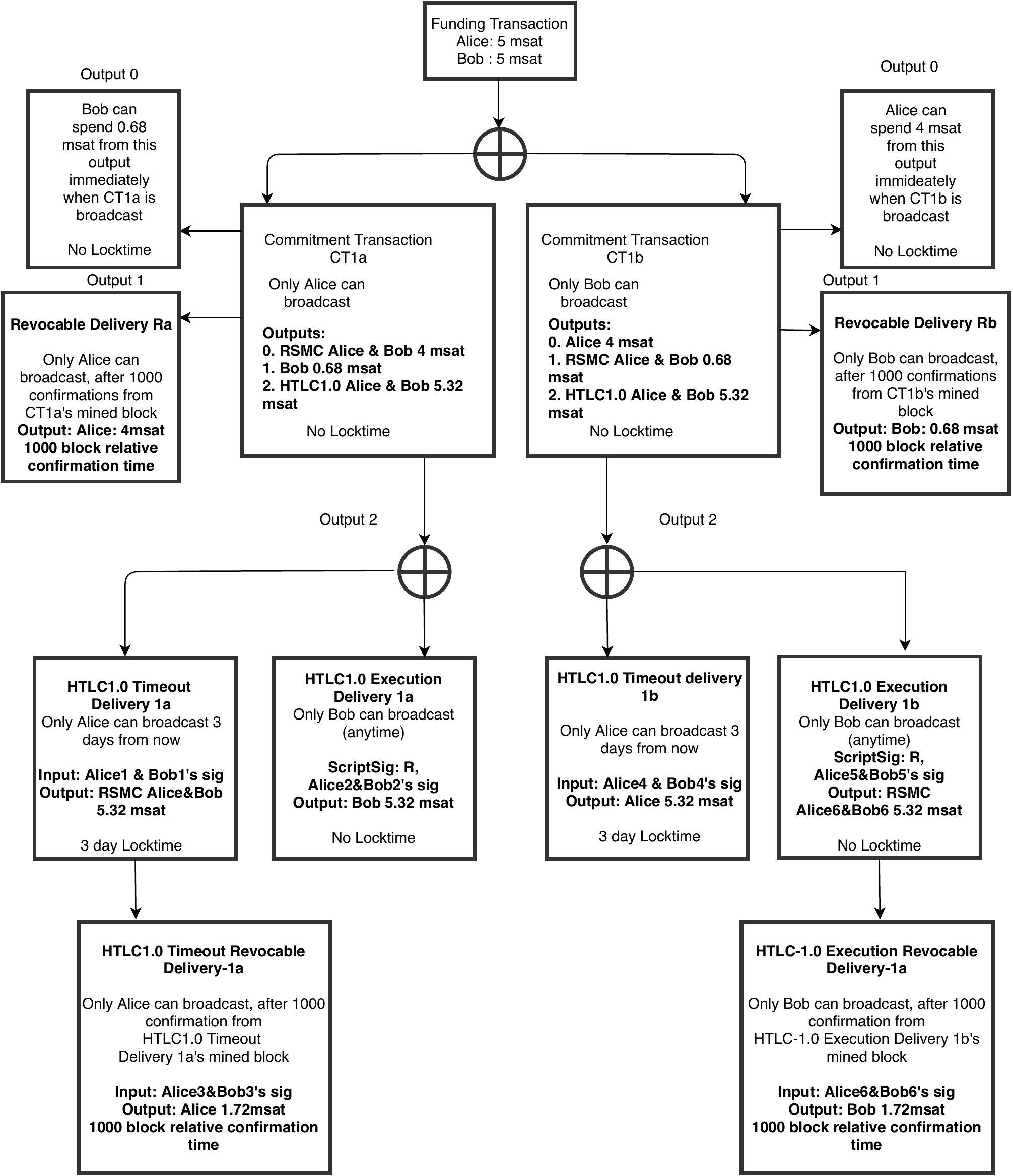}
    \caption{Revocable \textit{HTLC1.0}}
    \label{fig21}
\end{figure*}
Detailed construction of an instance of 2-party \emph{HTLC1.0} is explained with an example shown in Fig. \ref{21}. Alice has established an off-chain  contract with Bob for transferring 1 msat. Rate of griefing-penalty being 0.001 per minute. The terms of the contract is as follows:  \emph{Given $H=\mathcal{H}(x)$ in the contract, Bob can claim fund of 1 msat from Alice contingent to the knowledge of $x$, within a time-period of 3 days. If Bob fails to do so, then after a timeout of 3 days, it pays a penalty of 4.32 msat to Alice.}

The establishment of two-party \textit{HTLC1.0} between Alice and Bob has been illustrated in Fig. \ref{fig21}, the structure being similar to construction of \emph{Off-Chain Revocable HTLC} \cite{poon2016bitcoin}. Both parties have locked funds of 5 msat each, which gets included as the Funding Transaction. Bob locks 4.32 msat and Alice locks 1 msat into \textit{HTLC1.0}. Bob can withdraw the entire amount contingent to the knowledge of preimage corresponding to the payment hash. If Bob fails to respond, upon expiration of locktime Alice claims the entire amount. Thus both the parties mutually agree to form second commitment transaction (CT1a/CT1b). Output 2 of CT1a describes how funds get locked in \textit{HTLC1.0}. 5.32 msat will be encumbered in an \textit{HTLC1.0}. If a party wants to unilaterally close the channel then it broadcasts latest Commitment Transaction. The parties are remunerated as per terms of the contract. If CT1a is broadcasted and Bob has produced R within 3 days, it can immediately claim the fund of 5.32 msat by broadcasting \textit{HTLC1.0} Execution Delivery 1a. Revocable Sequence Maturity Contract (RSMC) embedding \cite{poon2016bitcoin} used in the output \textit{HTLC1.0} Timeout Delivery 1a ensures that if Alice broadcasts this transaction, it has to wait for 1000 block confirmation time before it can spend 5.32 msat. This extra waiting time serves as a buffer time for resolving dispute. If Alice had made a false claim of CT1a being the latest state of the channel, Bob will raise a dispute and spend Alice's channel deposit.

The same state of channel is replicated in CT1b. However, the difference lies in how each party can spend their respective output with respect to the copy of the transaction they have. If CT1b is broadcasted and Bob has not been able to produce R within a period of 3 days, then it can claim fund of 5.32 msat after 3 days by broadcasting \textit{HTLC1.0} Timeout delivery 1b. If Bob has the preimage R, it can immediately broadcast \textit{HTLC1.0} Execution Delivery 1b. However this output is encumbered by 1000 block confirmation time, the explanation being the same as we had stated for \textit{HTLC1.0} Timeout Delivery 1a. These changes can be easily integrated into the Bitcoin script.

\subsection{An instance of Multihop \textit{HTLC1.0}}
\begin{figure}[!ht]
    \centering
    \includegraphics[width=10cm]{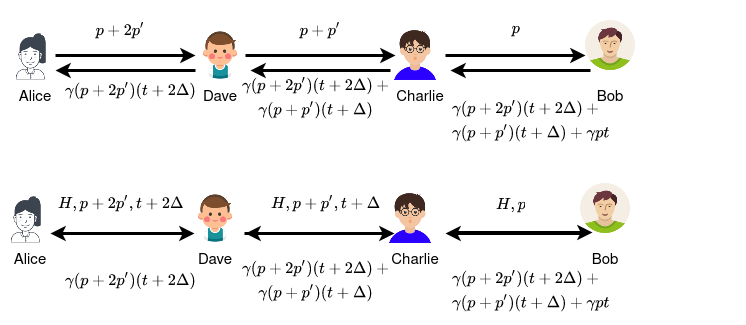}
    \caption{Formation of contract in \textit{HTLC1.0}}
    \label{fig12}
\end{figure}

We desribe the use of \emph{HTLC1.0} for multihop payment using an example. \emph{Alice} wants to transfer $p$ coins to \emph{Bob}. \emph{Bob} shares the hash $H$ with \emph{Alice} offline. This is used as the condition  for the off-chain contracts established in the route forwarding the payment. Given the rate of griefing-penalty as $\gamma$ per unit of time, $0\leq \gamma < 1$, and locktime of the contract being $(t+2\Delta)$, \emph{Dave} is expected to lock $\gamma (p+2p')(t+2\Delta)$ coins as griefing-penalty in the off-chain contract, $(p+2p')(t+2\Delta)$ being the collateral cost in channel \emph{Alice-Dave}. If \emph{Dave} provides the preimage of $H$ within this period, he will claim $p+2p'$ coins and withdraw $\gamma (p+2p')(t+2\Delta)$ coins locked in the contract. \emph{Dave} forwards a conditional payment of $p+p'$ coins to \emph{Charlie} by forming similar off-chain contract using payment hash $H$ and locktime $(t+\Delta)$. Upon griefing, \emph{Charlie} must pay a compensation of $\gamma (p+p')(t+\Delta)$. However, this amount is not sufficient to compensate both \emph{Dave} and \emph{Alice}. Hence he has to lock a cumulative griefing-penalty $\gamma (p+2p')(t+2\Delta)+\gamma (p+p')(t+\Delta)$ in the contract. This cumulative griefing-penalty is the summation of collateral cost in channel \emph{Alice-Dave} and \emph{Dave-Charlie}. \emph{Charlie} forwards a conditional payment of $p$ coins to \emph{Bob} by forming an off-chain contract for locktime of $t$ units. \emph{Bob} has to lock $\gamma (p+2p')(t+2\Delta)+\gamma (p+p')(t+\Delta)+\gamma p t$ coins. This amount is the cumulative penalty to be distributed among \emph{Alice, Dave} and \emph{Charlie}, if Bob griefs. The entire payment protocol construction is shown in Fig. \ref{fig12}.

Suppose \emph{Bob} griefs and refuses to release the preimage of $H$, waiting for time $t$ to elapse. He will pay a compensation of $\gamma (p+2p')(t+2\Delta)+\gamma (p+p')(t+\Delta)+\gamma p t$ coins to \emph{Charlie}, as per the terms of the contract. After the timelock $t$ expires, \emph{Charlie} goes on-chain. He closes the channel, unlocks $p$ coins and claims $\gamma (p+2p')(t+2\Delta)+\gamma (p+p')(t+\Delta)+\gamma p t$ coins as the compensation. He requests \emph{Dave} to cancel the off-chain contract offering a compensation of $\gamma (p+2p')(t+2\Delta)+\gamma (p+p')(t+\Delta)$. \emph{Dave} cancels the contract off-chain, unlocks $p+p'$ coins from the contract and claims the compensation from \emph{Charlie}. If \emph{Charlie} decides to grief, \emph{Dave} can claim the compensation by going on-chain and closing the channel. \emph{Dave} requests \emph{Alice} to cancel the contract by offering a compensation of $\gamma (p+2p')(t+2\Delta)$. Thus except \emph{Bob}, none of the parties lose funds in order to compensate any of the affected parties. 

\subsubsection{Problem of Reverse-Griefing in HTLC1.0}

\label{revrgrief}
\begin{figure}[!ht]
    \centering
    \includegraphics[width=10cm]{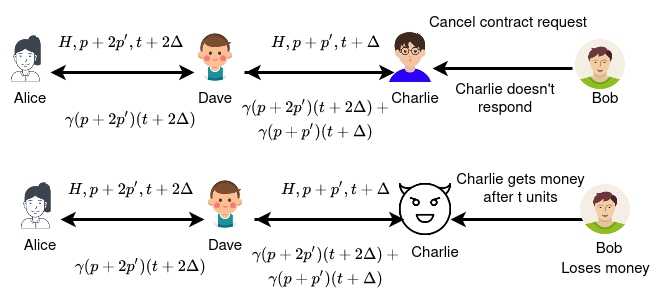}
    \caption{Reverse-Griefing attack by \emph{Charlie}}
    \label{32}
\end{figure}

A major drawback of the protocol is that with the introduction
of griefing-penalty, a malicious party can now ascribe the
blame of griefing on an honest party as well. In the previous
example, if \emph{Alice} uses a wrong hash value or the HTLC timeout period is closer to the current block height of blockchain or the value of transaction forwarded is less than agreed value \cite{error}, then \emph{Bob} would cancel the payment. However, \emph{Charlie} can deny settling
the contract off-chain. \emph{Bob} has no way to prove his innocence.
Ultimately, with elapse of locktime, \emph{Charlie} goes on-chain, claiming
\emph{Bob's} money, as shown in Fig. \ref{32}. This attack is termed as \emph{Reverse Griefing}. 


Though there is problem with this construction, it is quite simple and requires minimal amount of changes to the existing \emph{HTLC protocol} for the purpose of implementation in Lightning Network. We discuss why an honest rational party may not be easily motivated to mount reverse-griefing attack: 

\begin{itemize}
\item The attacker needs to wait for the entire locktime before it can collect penalty from the counterparty. In a path of length $n$, locktime of the contract established by the adversary can range from $\Delta$ units to $n\Delta$ units, depending upon its position in the path.
\item The attacker needs to go on-chain with the \textit{HTLC1.0} to redeem the penalty and pay the mining fee. If the cumulative penalty earned by the attacker is less than the bitcoin transaction fee, reverse-griefing is a loss making strategy. 
\item Since reverse-griefing affects a single honest node, the adversary's intention of blocking funds across the network becomes more costly as it needs to coordinate and mount several such attacks. If it wants to avoid paying the griefing-penalty, it has to corrupt more than one nodes in the path and devise a strategy accordingly.

\item In case the intention was to block the counterparty's funds, the attack fails to block any of the other edges that the counterparty has with other honest nodes. 
\item In case the attacker has a high degree centrality or betweenness centrality, it is even more foolish to mount a reverse-griefing attack as it is probable that the expected processing fee for the duration of the locktime might be far greater than the cumulative penalty reward. 
\item The rate of griefing-penalty can be lowered in order to disincentivize a party from reverse-griefing. 
\end{itemize}


%
%

However, there exist several nodes in the network which earn very low processing fee during their entire channel lifetime. Either they charge very negligible amount of fee for large valued transaction \cite{beres2019cryptoeconomic} or they remain inactive for most of their lifetime in the network. Such nodes have higher tendency to deviate as the profit earned by reverse-griefing is higher than the total anticipated processing fee. Acceptability of \textit{HTLC1.0} is hence subjected to such arbitrary behavior.

\section{Our Proposed Protocol using Griefing Penalty}
\label{rg1}
It is observed in \emph{HTLC1.0} that establishing a single contract with minimal changes to the script is not sufficient to protect a party from being cheated.  To avoid the problem of reverse-griefing, we propose a new payment protocol for Lightning Network, termed as Hashed Timelock Contract with Griefing-Penalty or \textit{HTLC-GP}. 

In this protocol, locking of penalty and locking of the transaction amount must be executed in separate rounds. Instead of both parties locking their funds into a single contract, the payer locks fund in one contract, the \emph{Payment Contract}, and the payee locks his penalty in a separate contract, the \emph{Cancellation Contract}. The two contracts are bound together using two distinct hashes, termed as \emph{Payment Hash} and \emph{Cancellation Hash}. The payee can unlock the penalty deposited in \emph{Cancellation Contract} either by providing the preimage to the first hash, i.e. the payment hash, or by providing the preimage to the second hash, i.e. the cancellation hash. The problem with this arrangement is the order in which the contracts must be established. If the payment contract gets established first then the payee can still grief without establishing the cancellation contract. In this way it can avoid any payment of griefing penalty. Thus we put the payer at an advantage by asking the payee to lock penalty into the \emph{Cancellation Contract} and forward it to the payer in the first round. After the payer receives the contract, it will lock the fund into the \emph{Payment Contract} and forward it to the payee. Since the payee is in possession of preimages corresponding to both the hashes, even if the payer denies forming the payment contract, it cannot mount a reverse-griefing attack on payee. After a certain time, the payee will cancel the contract by releasing the cancellation hash. 

\subsubsection{Construction of 2 party HTLC-GP}
\label{2party}
Consider an example where Alice and Bob have locked funds of 5 msat each, which gets included as the Funding Transaction. Alice intends to transfer 1 msat to Bob. Rate of griefing-penalty being 0.001 per minute. The terms of the contract is as follows:  \emph{Given $H=\mathcal{H}(x)$ in the contract, Bob can claim fund of 1 msat from Alice contingent to the knowledge of $x$, within a time-period of 3 days. If Bob fails to do so, then after a timeout of 3 days, it pays a penalty of 4.32 msat to Alice.}
  \begin{figure*}[!ht]
    \centering
    \includegraphics[scale=0.6]{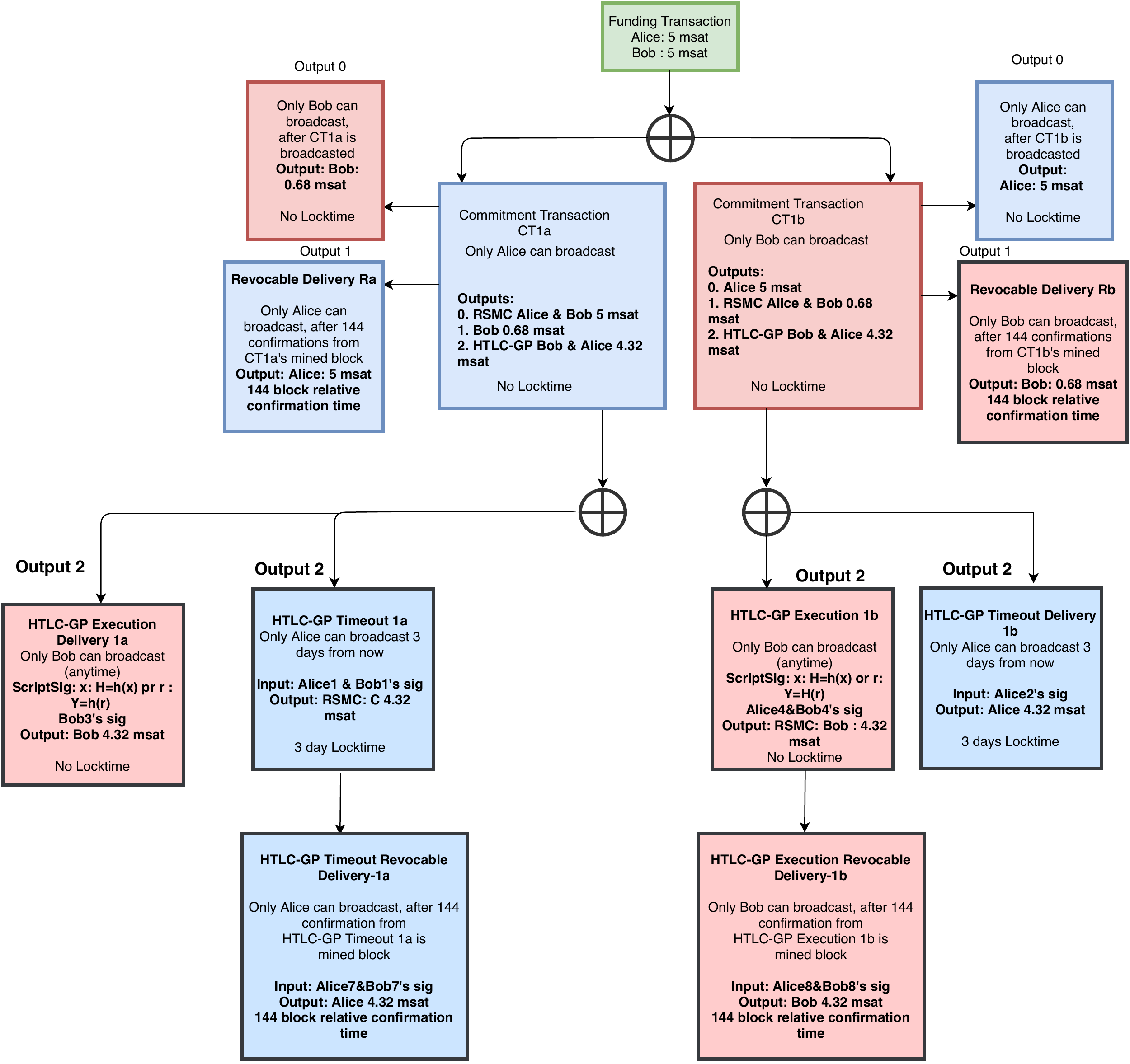}
    \caption{Revocable \emph{HTLC-GP} using two hashes: First Round Locking}
    \label{new1}
\end{figure*}

\begin{figure*}[!ht]
    \centering
    \includegraphics[scale=0.5,angle=90]{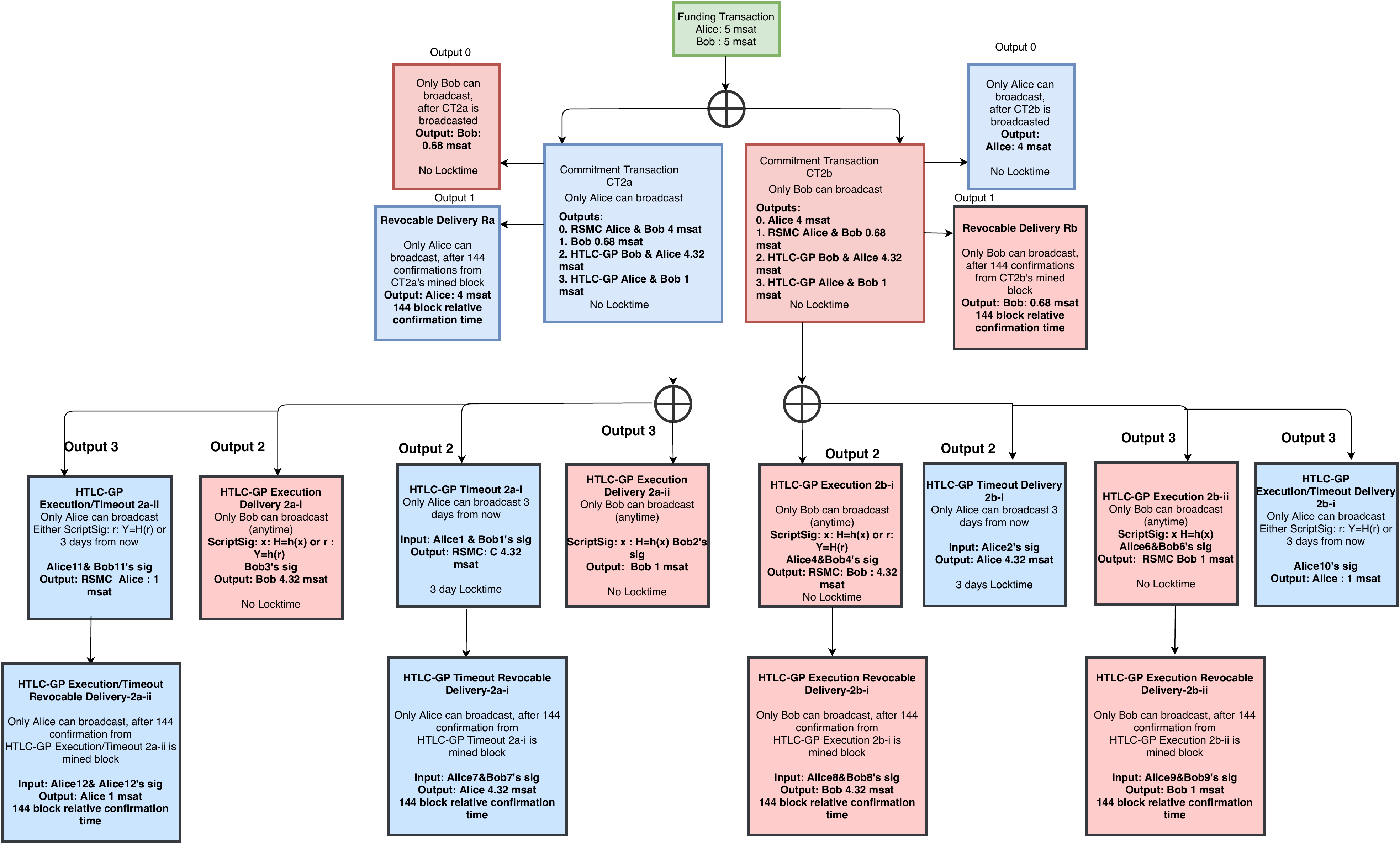}
    \caption{Revocable HTLC-GP using two hashes: Second Round Locking}
    \label{newconstruct}
\end{figure*}

In the first round of Locking, Bob forwards the \emph{cancellation contract} to Alice by locking 4.32 msat. Bob can withdraw the entire amount contingent to the knowledge of preimage corresponding to the cancellation hash $Y$ or payment hash $H$. If Bob fails to respond, upon expiration of locktime Alice claims the entire amount. Thus both the parties mutually agree to form the commitment transaction (CT1a/CT1b) as shown in Fig. \ref{new1}.
The state of the channel is: Alice has balance of 5 msat, Bob has balance of 0.68 msat, money locked in \emph{HTLC-GP} is 4.32 msat. Output 2 of CT1a describes how funds get locked in HTLC-GP. 4.32 msat will be encumbered in an \emph{HTLC-GP}. If a party wants to unilaterally close the channel then it broadcasts latest Commitment Transaction. The parties are remunerated as per terms of the contract. If CT1a is broadcasted and Bob has produced $r$ or $x$ within 3 days, it can immediately claim the fund of 4.32 msat by broadcasting \emph{HTLC-GP} Execution Delivery 1a. Revocable Sequence Maturity Contract (RSMC) embedding \cite{poon2016bitcoin} used in the output \emph{HTLC-GP} Timeout Delivery 1a ensures that if Alice broadcasts this transaction after 3 days, it has to wait for 144 block relative confirmation time before it can spend 4.32 msat. This extra waiting time serves as a buffer time for resolving dispute. If Alice had made a false claim of CT1a being the latest state of the channel, Bob will raise a dispute and spend Alice's channel deposit.

The same state of channel is replicated in CT1b. However, the difference lies in how each party can spend their respective output with respect to the copy of the transaction they have. If CT1b is broadcasted and Bob has not been able to produce either $r$ or $x$ within a period of 3 days, then it can claim fund of 4.32 msat after 3 days by broadcasting \emph{HTLC-GP} Timeout delivery 1b. Bob's transaction \emph{HTLC-GP} Execution Delivery 1b is encumbered by 144 block relative confirmation time, the explanation being the same as we had stated for \emph{HTLC-GP} Timeout Delivery 1a.

In the second round of Locking, Alice and Bob update the state of the channel and create new commitment transaction CT2a/CT2b, as shown in Fig.\ref{newconstruct}. The previous \emph{HTLC-GP} based on locking of penalty, is carried forward as it is still unresolved. In this round, Alice forwards the \emph{payment contract} to Bob by locking 1 msat. The state of the channel is: Alice has balance of 4 msat, Bob has balance of 0.68 msat, money locked in first \emph{HTLC-GP} is 4.32 msat and money locked in second \emph{HTLC-GP} is 1 msat. Bob can claim the money from both the contracts contingent to the knowledge of preimage corresponding to the payment hash $H$. If Bob reveals the preimage corresponding to cancellation hash, Alice withdraws 1 msat and Bob withdraws 4.32 msat from the contract. If Bob doesn't respond before the locktime expires, Alice claims the money locked in both the contracts. 

\subsubsection*{HTLC-GP Script} 
The structure of the script is as per the convention used in \cite{lnscript}. For implementation of HTLC-GP in Lightning Network, we discuss how to design the output scripts for \emph{Cancellation Contract} and \emph{Payment Contract}.
\label{script}
\vspace{0.2cm}\\
\underline{HTLC-GP Offered Cancellation Contract}: Bob offers this script to Alice. This output sends funds to either the remote node after the HTLC-GP timeout or using the revocation key, or to an HTLC-GP -success transaction either with a successful payment preimage or cancellation preimage. The output is a P2WSH, with a witness script:
\begin{itemize}[leftmargin=*]
\item Release the funds if the script is signed by the revocation key (\emph{revocationpubkey}).
\item If the above condition fails, then check if HTLC-GP public key of the party not publishing the commitment (remote public key), i.e. of Alice, was provided. Now check which of the condition holds true:
\begin{itemize}[leftmargin=*]
\item The publisher of the commitment, i.e. Bob, can publish the HTLC-GP-success by using the notif clause. It ignores the condition when the remote public key is not provided. HTLC-GP-success condition can be realized if either of the condition is satisfied:
\begin{itemize}[leftmargin=*]
\item Bob can use the preimage of \emph{cancellation hash}. Release the funds if the preimage is released and signed by both Alice and Bob,
\item Bob can use the preimage of \emph{payment hash}. Release the funds if the preimage is released and signed by both Alice and Bob.
\end{itemize}
\item If Bob didn't react, Alice can publish the HTLC-GP-timeout transaction.
\end{itemize}
\end{itemize}
The Bitcoin script structure is shown in Fig.\ref{script1}.

\begin{figure*}[!ht]
\caption{Script Structure: Offered Cancellation Contract}
\label{script1}
\begin{tcolorbox}[colback=black!5,colframe=black!75!black,title={} ]
 \texttt{OP\_DUP OP\_HASH160 $\langle$ RIPEMD160 ( SHA256 ( revocationpubkey ))$\rangle$ OP\_EQUAL}\\
\texttt{OP\_IF}\\
\hspace*{1.5em}\texttt{OP\_CHECKSIG}\\
\texttt{OP\_ELSE}\\
\hspace*{1.5em}\texttt{$\langle$ remote\_htlcgppubkey$\rangle$ OP\_SWAP OP\_SIZE 32 OP\_EQUAL}\\
\hspace*{1.5em}\texttt{OP\_NOTIF}\\
\hspace*{1.5em}\hspace*{1.5em}\texttt{OP\_IF}\\
\hspace*{1.5em}\hspace*{1.5em}\hspace*{1.5em}\texttt{OP\_HASH160 $\langle$ RIPEMD160 ( payment\_hash )$\rangle$ OP\_EQUALVERIFY}\\
\hspace*{1.5em}\hspace*{1.5em}\hspace*{1.5em}\texttt{2 OP\_SWAP $\langle$ local\_htlcgppubkey $\rangle$ 2 OP\_CHECKMULTISIG}\\
\hspace*{1.5em}\hspace*{1.5em}\texttt{OP\_ELSE}\\
\hspace*{1.5em}\hspace*{1.5em}\hspace*{1.5em}\texttt{OP\_HASH160 $\langle$ RIPEMD160 ( cancellation\_hash) $\rangle$ OP\_EQUALVERIFY} \\
\hspace*{1.5em}\hspace*{1.5em}\hspace*{1.5em}\texttt{2 OP\_SWAP $\langle$ local\_htlcgppubkey $\rangle$ 2 OP\_CHECKMULTISIG}\\
\hspace*{1.5em}\hspace*{1.5em}\texttt{OP\_ENDIF}\\
\hspace*{1.5em}\texttt{OP\_ELSE}\\
\hspace*{1.5em}\hspace*{1.5em}\texttt{OP\_DROP $\langle$ cltv\_expiry $\rangle$ OP\_CHECKLOCKTIMEVERIFY OP\_DROP}\\
\hspace*{1.5em}\hspace*{1.5em}\texttt{OP\_CHECKSIG}\\
\hspace*{1.5em}\texttt{OP\_ENDIF}\\
\texttt{OP\_ENDIF}
\end{tcolorbox}
\end{figure*}
\vspace*{0.2cm}
\underline{HTLC-GP Offered Payment Contract}:  Alice offers this script to Bob. This output sends funds to either an HTLC-timeout transaction after the HTLC-timeout or to the remote node using either the payment preimage or cancellation image or the revocation key. The output is a P2WSH, with a witness script:
\begin{itemize}[leftmargin=*]
\item Release the funds if the script is signed by the revocation key (\emph{revocationpubkey}).
\item If the above condition fails, then check if HTLC-GP public key of the party not publishing the commitment (remote public key), i.e. of Bob, was provided. Now check which of the condition holds true:
\begin{itemize}[leftmargin=*]
\item The publisher of the commitment, i.e. Alice, can publish the HTLC-GP-timeout by using the notif clause. 
\item Else, Bob can publish HTLC-GP success if any of the condition holds true:
\begin{itemize}[leftmargin=*]
\item Bob can use the preimage of \emph{cancellation hash}. Release the funds if the preimage is released and signed by both Alice and Bob,
\item Bob can use the preimage of \emph{payment hash}. Release the funds if the preimage is released and signed by both Alice and Bob.
\end{itemize}
\end{itemize}
\end{itemize}
The Bitcoin script structure is shown in Fig.\ref{script2}.

\begin{figure*}[!ht]
\caption{Script Structure: Offered Payment Contract}
\label{script2}
\begin{tcolorbox}[colback=black!5,colframe=black!75!black,title={} ]
\texttt{OP\_DUP OP\_HASH160 $\langle$ RIPEMD160 ( SHA256 ( revocationpubkey ))$\rangle$ OP\_EQUAL}\\
\texttt{OP\_IF}\\
\hspace*{1.5em}\texttt{OP\_CHECKSIG}\\
\texttt{OP\_ELSE}\\
\hspace*{1.5em}\texttt{$\langle$ remote\_htlcgppubkey$\rangle$ OP\_SWAP OP\_SIZE 32 OP\_EQUAL}\\
\hspace*{1.5em}\texttt{OP\_NOTIF}\\
\hspace*{1.5em}\hspace*{1.5em}\texttt{OP\_DROP 2 OP\_SWAP $\langle$ local\_htlcgppubkey $\rangle$ 2 OP\_CHECKMULTISIG}\\
\hspace*{1.5em}\texttt{OP\_ELSE}\\
\hspace*{1.5em}\hspace*{1.5em}\texttt{OP\_IF}\\
\hspace*{1.5em}\hspace*{1.5em}\hspace*{1.5em}\texttt{OP\_HASH160 $\langle$ RIPEMD160 ( cancellation\_hash) $\rangle$ OP\_EQUALVERIFY} \\
\hspace*{1.5em}\hspace*{1.5em}\hspace*{1.5em}\texttt{OP\_CHECKSIG}\\
\hspace*{1.5em}\hspace*{1.5em}\texttt{OP\_ELSE}\\
\hspace*{1.5em}\hspace*{1.5em}\hspace*{1.5em}\texttt{OP\_HASH160 $\langle$ RIPEMD160 ( payment\_hash )$\rangle$ OP\_EQUALVERIFY}\\
\hspace*{1.5em}\hspace*{1.5em}\hspace*{1.5em}\texttt{OP\_CHECKSIG}\\
\hspace*{1.5em}\hspace*{1.5em}\texttt{OP\_ENDIF}\\
\hspace*{1.5em}\texttt{OP\_ENDIF}\\
\texttt{OP\_ENDIF}
\end{tcolorbox}
\end{figure*}

In the next section, we provide an instantiation of multihop payment using \emph{HTLC-GP}.

\section{Multihop Payment using HTLC-GP}
\label{paygp}
\subsection{System Model}
\label{sys}


  The topology of the Lightning Network is known by all the participants in the network since opening or closing of a payment channel is recorded on the blockchain. Pairs of honest users, sharing a payment channel, communicate through secure and authenticated channels. An honest party willing to send funds to another party, will adhere to the protocol. It will not tamper with the terms and conditions of off-chain contract meant for each of the intermediate payment channels, involved in relaying the payment. The model of communication is considered to be synchronous, with all the parties following a global clock for settling payments off-chain. We assume that all the nodes in the network are rational, whose intention is to maximize the profit earned.

\subsection{Objective} 
\begin{itemize}

\item \textbf{Guaranteed compensation for an honest node}: All the honest parties affected by the griefing attack will be remunerated by the griefer. Except the griefer, none of them must lose fund in order to pay compensation to any of the affected parties. 
\item \textbf{Payer and Payee's Privacy}: None of the intermediate nodes involved in routing a payment must be able to identify its exact position in the path as well as figure out the identities of sender and receiver of payment. 

\end{itemize}

\subsection{Adversarial Model \& Assumptions}
\label{adv1}
An adversary introduces multiple Sybil nodes and places them strategically in the network in order to maximize the collateral damage. Such Sybil nodes may be involved in self-payment or transfer funds from one Sybil node to the other for mounting griefing attack. The Sybil nodes may also act as intermediate nodes in a path of payment. The adversary can perform the following arbitrary actions in order to keep funds locked in the network for a substantial amount of time:
\begin{itemize}
\item  It withholds the solution without resolving the incoming off-chain payment request.
\item It may refrain from forwarding the off-chain payment request to the next neighbour.
\item It just refuses to sign any incoming contract request. 
\end{itemize}
We assume that in a path executing the payment, at least one node will be honest. Thus in the worst case, except one node (either sender or receiver or any intermediate party), rest all the parties may be corrupted and controlled by the adversary. We also assume that an honest party can cannot be denied going on-chain by the adversary during the protocol. Using untrusted/semi-trusted third party service provider, \emph{WatchTowers}, prevents such attackers from mounting \emph{time dilation attacks} \cite{riard2020time} and censoring transactions.

\subsection{Our proposed Construction}
\label{multigp}
Given an instance of Lightning Network, for secure transfers of funds from sender $U_0$ to the receiver, the former selects an optimal route for transferring funds to the payee, as per its routing strategy. Since the path length $n$, we index the receiver as $U_n$. Let the path be $P=\langle U_0,U_1,\ldots,U_n\rangle$, via which payer $U_0$ will relay fund of value $\alpha$ to payee $U_n$, each $U_i$ is a node in the graph and $(U_{i},U_{i+1}), i \in [0,n-1]$ denotes a payment channel. Each party $U_i, i \in [1,n-1]$ charge a service fee of $fee(U_i)$ for relaying the fund. Hence the total amount that $U_0$ needs to transfer is $\tilde{\alpha}=\alpha+\Sigma_{i=1}^{n-1} fee(U_i)$. We denote each $\alpha_i=\tilde{\alpha}-\Sigma_{j=1}^i fee(U_j), i \in [1,n-1],  \alpha_0=\tilde{\alpha}$ and $\alpha_{n-1}=\alpha$. Each node $U_i$ samples pair of secret key and public key $(sk_i,pk_i)$, the public key of each node is used to encrypt the information of establishing contract with the neighbouring node. 

\subsubsection*{Parameters used}
\begin{itemize}[leftmargin=*]

\item \textbf{Rate of Griefing-penalty $\gamma$}: It decides the amount to be deducted per unit time as compensation from the balance of a node responsible for griefing. It is set as a system parameter with $0 \leq \gamma \leq 1$, measured in terms of per minute. 

\item \textbf{Routing Attempt Cost $\psi$}: $U_0$ has to figure out the path by probing channels that will be able to route the transaction. This may require several attempts and hence adds an extra computational as well as resource overhead on $U_0$. Thus $U_0$ adds the cost of routing attempt to the compensation withdrawn from griefer. This is a variable quantity and the quantity is kept hidden from other nodes but generally, a sender sets the value $\psi t_{0} \geq  \alpha ((k+1) t_0+\Sigma_{l=1}^{k} l\Delta), k \in \mathbb{N}$ and preferably $ k> 3$. Here $k$ is the masking factor, $t_0$ is the time period of the contract established between $U_0$ and $U_1$ and $\Delta$ is the time taken for a transaction to settle on-chain.

\end{itemize}
\subsubsection*{Computing Griefing-Penalty} $U_0$ shares $\phi(n)$ with $U_n$, where $\phi$ is a function used for blinding the exact value of $n,\  \phi(n).\alpha t_{n-1} \approx ((\psi + \alpha_{0})t_{0}+\Sigma_{j=1}^{n-1}\alpha_{j}t_{j}) $, adding the extra cost for routing to the compensation it must claim from $U_1$. Similar to HTLC, $t_{n-1}$ is the least timeout period, assigned to the off-chain contract between $U_{n-1}$ and $U_n$ where $t_{n-1}>\Delta$. For rest of the off-chain contracts established between $U_i$ and $U_{i+1}, i \in [0,n-2]$, the timeperiod of the contract $t_{i}>t_{i+1}+\Delta$. Adding the routing attempt costs to the compensation disallows $U_1$ from inferring the identity of the sender of a particular payment. It cannot distinguish whether $\psi$ is routing attempt fee or the cumulative flow from any predecessors of $U_0$.  Even other nodes must not be able to figure out their position in the path with the information of cumulative griefing-penalty. The proof has been discussed in \emph{Theorem 2}, in Section \ref{secgp1}.

The maximum compensation which can be earned by $U_i, i \in [1,n-1]$ is $\gamma.\alpha_i t_i$, where $\alpha_i$ is the amount to be transferred to $U_{i+1}$, if the contract is resolved successfully within $t_i$. If $U_{i+1}$ is at fault, then it has to pay compensation to all the parties which got affected starting from $U_i$ till $U_0$. Hence compensation charged by each channel $(U_k,U_{k+1}), k \in [0,i]$, must be withdrawn from the faulty node $U_{i+1}$. The total griefing-penalty to be paid is $\gamma.(\Sigma_{j=1}^{i} (\alpha_j t_j)+(\alpha_0+\psi)t_0)$, so that each party $U_m, m \in [1,i]$, gets a compensation of $\gamma.\alpha_m t_m$ and $U_0$ withdraws a compensation of $\gamma (\psi+\tilde{\alpha})t_0$.

\subsubsection{Protocol Description} 
\label{pd}

 Our protocol involves the following three phases: 

\noindent \texttt{\underline{Pre-processing Phase}} 
\begin{itemize}[leftmargin=*]
\item $U_n$ samples the preimages $x$ and $r$, $x\neq r$ and constructs the two hashes: $H=\mathcal{H}(x)$ and $Y=\mathcal{H}(r)$. \item It shares $H,Y$ with the payer, $U_0$. The payer uses standard onion routing \cite{goldschlag1999onion} for propagating the information needed by each node $U_i, i \in [1,n]$, across the path $P$. 
\item The cumulative griefing-penalty for node $U_0$ is defined as tgp$_{0}=\gamma (\psi+\tilde{\alpha})t_0$ and for any node $U_i, i\in [1,n-1]$ as tgp$_{i}= \gamma.(\Sigma_{j=1}^{i} (\alpha_j t_j)+(\alpha_0+\psi)t_0)$.

\item $U_0$ sends $M_0=E(\ldots E(E(E(\phi,Z_n,pk_{n}),Z_{n-1},pk_{n-1}), Z_{n-2},pk_{n-2})$\ldots$,Z_{1},pk_{1})$ to $U_1$, where $Z_i=(H,Y,\alpha_i,t_{i-1},\\\textrm{tgp}_{i-1},U_{i+1}), i \in [1,n-1]$ and $Z_n=(H,Y,\alpha_{n-1},t_{n-1}, \textrm{tgp}_{n-1},null)$. Here $M_{i-1}=E(M_{i},Z_{i},pk_{i})$ is the encryption of the message $M_{i}$ and $Z_{i}$ using public key $pk_{i}$, $M_n=\phi$. 
\item $U_1$ decrypts $M_0$, gets $Z_1$ and $M_1$. $M_1=E(\ldots E(E(E(\phi,Z_n,pk_{n}),Z_{n-1},pk_{n-1}), Z_{n-2},pk_{n-2}),\ldots,Z_{2},pk_{2})$ is forwarded to the next destination $U_2$. This continues till party $U_n$ gets $E(\phi,Z_n,pk_n)$.  
\end{itemize}
%

\noindent \texttt{\underline{Two-Round Locking Phase}}
It involves two rounds: \emph{establishing Cancellation Contract} and \emph{establishing Payment Contract}.
\begin{itemize}[leftmargin=*]
\item \textit{Establishing Cancellation Contract}: Since the flow of griefing-penalty is in the opposite direction of the actual payment, it is logical for $U_n$ to initiate this round. 
\begin{itemize}[leftmargin=*]
\item $U_n$ decrypts to get $Z_n$. It checks $\gamma \phi(n)t_{n-1}\stackrel{?}{\approx}tgp_{n-1}$ and $\alpha_{n-1}\stackrel{?}{=}\alpha$. If this holds true, it forms a contract with $U_{n-1}$, locking $tgp_{n-1}$. 
\item For the rest of the parties, $U_{i}, i \in [1,n-1]$ first checks $tgp_{i}-\gamma \alpha_{i} t_{i} \stackrel{?}{=} tgp_{i-1}$ and then forms the off-chain contract with $U_{i-1}$, locking $tgp_{i-1}$. 
\item The terms of the contract is defined as follows: \emph{`$U_{i+1}$ can withdraw the amount $tgp_i=\gamma.(\Sigma_{j=1}^{i} (\alpha_j t_j)+(\alpha_0+\psi)t_0)$ from the contract provided it reveals either $x:H=\mathcal{H}(x)$ or $\ r: Y=\mathcal{H}(r)$ within a period of $t_i$ else $U_{i}$ claims this amount as griefing-penalty after the elapse of the locktime.'}.  
\end{itemize}
\emph{Bad Case}: If $U_{i-1}$ denies signing the cancellation contract then $U_i$ will abort. However, $U_n$ locks a substantial amount as griefing-penalty, thus it will wait for a bounded amount of time and withdraw the money, thereby aborting from the process. We denote this time as $\delta: \delta\leq t_{n-1}$. If $U_{n-1}$ stops responding after establishment of the cancellation contract, $U_n$ releases on-chain the preimage $r$ corresponding to cancellation hash after waiting for $\delta$ units of time and unlock the penalty $tgp_{n-1}$ from the contract. The preimage $r$ is now used by other parties $U_j,j \in [i+1,n]$ to cancel their respective off-chain contracts with $U_{j-1}$. So even if $U_i$ aborts, $U_{i+1}$ can go on-chain, close the channel and withdraw the amount locked in the contract. 

The pseudocode of the first round of Locking Phase for $U_n$, any intermediate party $U_i, i \in [1,n-1]$ and payer $U_0$ is stated in Procedure \ref{algo:lock}, Procedure \ref{algo:lock1} and Procedure \ref{algo:lock2} respectively.


\begin{proc}[H]
    \SetKwInOut{Input}{Input}
    \SetKwInOut{Output}{Output}

    \caption{Establishing Cancellation Contract: First Round of Locking Phase for $U_n$ }
        \label{algo:lock}
        
\textbf{Input}: $(Z_{n},\phi(n),\gamma, \alpha)$  \\
$U_n$ parses $Z_{n}$ and gets $H', Y',\alpha',t',\textrm{tgp}_{n-1}$.\\
\If{$t'\geq t_{now}+\Delta$ and $\alpha'\stackrel{?}{=}\alpha$ and $\gamma(\phi(n)\alpha)t'\approx \textrm{tgp}_{n-1}$ and $H'\stackrel{?}{=}H$ and $Y'\stackrel{?}{=}Y$  and $remain(U_{n},U_{n-1})\geq \textrm{tgp}_{n-1}$}
{

    Send $\textrm{Cancel\_Contract\_Request}(H,Y,t',\textrm{tgp}_{n-1},\gamma)$ to $U_{n-1}$\\
     \If{acknowledgement received from $U_{n-1}$}
{      
  $remain(U_{n},U_{n-1})=remain(U_n,U_{n-1})-\textrm{tgp}_{n-1}$\\
  establish $Cancel\_Contract(H,Y,t',\textrm{tgp}_{n-1})$ with $U_{n-1}$\\
     Record $t_{n}^{form}=current\_clock\_time$\\
     }
     \Else
     {
       abort
     }

 }
 \Else
{

 abort.
}
 
 \end{proc}

\begin{proc}[H]
    \SetKwInOut{Input}{Input}
    \SetKwInOut{Output}{Output}

    \caption{Establishing Cancellation Contract: First Round of Locking Phase for $U_i, i\in [1,n-1]$ }
        \label{algo:lock1}
        
\textbf{Input}: $(H',Y',t',\textrm{tgp}_{i},\gamma)$  \\
$U_i$ parses $Z_{i}$ and gets $H, Y,\alpha_{i},t_{i-1},\textrm{tgp}_{i-1}$.\\

\If{$H'\stackrel{?}{=}H$ and $Y\stackrel{?}{=}Y'$ and $t'+\Delta\stackrel{?}{\leq} t_{i-1}$ and $\textrm{tgp}_{i}-\gamma \alpha_i t'\stackrel{?}{=}\textrm{tgp}_{i-1}$ and $remain(U_i,U_{i+1})\geq \alpha_i$ and $remain(U_{i},U_{i-1})\geq \textrm{tgp}_{i-1}$ and (current\_time not close to contract expiration time)}
{

  Sends acknowledgement to $U_{i+1}$ and wait for the off-chain contract to be established\\
  
       Send $\textrm{Cancel\_Contract\_Request}(H,Y,t_{i-1},\textrm{tgp}_{i-1},\gamma)$ to $U_{i-1}$\\

     \If{acknowledgement received from $U_{i-1}$}
{      
  $remain(U_{i},U_{i-1})=remain(U_i,U_{i-1})-\textrm{tgp}_{i-1}$\\
  establish $Cancel\_Contract(H,Y,t_{i-1},\textrm{tgp}_{i-1})$ with $U_{i-1}$\\
     }
     \Else
     {
     
       abort
     }

}
\Else
{

 abort.
}
    
\end{proc}

\begin{proc}[H]
    \SetKwInOut{Input}{Input}
    \SetKwInOut{Output}{Output}

    \caption{Establishing Cancellation Contract: First Round of Locking Phase for $U_0$ }
        \label{algo:lock2}
        
\textbf{Input}: $(H',Y',t',\textrm{tgp}',\gamma)$  \\

\If{$t'\stackrel{?}{=}t_0$ and $\textrm{tgp}'\stackrel{?}{=}\textrm{tgp}_0$ and $H'\stackrel{?}{=}H$ and $Y'\stackrel{?}{=}Y$  and $remain(U_{0},U_{1})\geq \alpha_0$}
{

  Sends acknowledgement to $U_{1}$\\
  Confirm formation of penalty contract with $U_1$\\
  Initiate the second round, establishment of payment contract\\

 }
 \Else
{

 abort.
}
 
 \end{proc}

\item \textit{Establishing Payment Contract}: $U_0$, upon receiving the cancellation contract, initiates the next round by establishing chain of contracts in the forward direction, till it reaches the payer $U_n$. This proceeds as normal \emph{HTLC}. 
\begin{itemize}[leftmargin=*]
\item Each node $U_i, i \in [0,n-1]$ forwards the terms of off-chain contract to $U_{i+1}$, locking $\alpha_i$.
\item  The off-chain contract is defined as follows: \emph{`$U_{i+1}$ can claim the amount $\alpha_i$ provided it reveals $x:H=\mathcal{H}(x)$ within a period of $t_i$. If not,  then $U_{i}$ withdraws the amount either contingent to the knowledge of $r:Y=\mathcal{H}(r)$ or after the elapse of locktime.'}
\end{itemize}
\end{itemize}

\emph{Bad Case}: If $U_{i+1}$ doesn't sign the payment contract, $U_i$ aborts from the process. Similar to the first round of locking phase, if $U_{n-1}$ doesn't form the payment contract within time $\delta$, $U_n$ releases the preimage $r$ and unlocks the penalty $tgp_{n-1}$ from the contract. $U_{i}$ will not be able to reverse-grief $U_{i+1}$ by aborting since the latter can go on-chain and withdraw the amount locked in the contract. 

The pseudocode of the second round of Locking Phase for $U_0$ and any intermediate party $U_i, i \in [1,n-1]$ is stated in Procedure \ref{algo:lock3} and Procedure \ref{algo:lock4} respectively.

\noindent \texttt{\underline{Release Phase}}: 
$U_n$ waits for $\delta$ units of time before initiating this round. If the payment contract received from $U_{n-1}$ is correct, it releases the preimage $x$ or payment witness and resolves the contract off-chain. If $U_{n-1}$ has not responded with incoming payment contract request or it has encountered any error (wrong payment or penalty value, invalid locktime) in the terms of incoming contract, it releases the cancellation preimage $r$. In case of dispute, it goes on-chain using either of the preimage for settling the contract. This is repeated for other parties $U_i, i \in [1,n-1]$, which upon obtaining the preimage claims payment from the counterparty or withdraws funds from the contract. 

If $U_{i+1}$ griefs and refuses to release preimage to $U_i$, it has to pay the a griefing-penalty for affecting the nodes $U_k, 0\leq k \leq i$, so that all the nodes obtain their due compensation. The pseudocode of the Release Phase for $U_n$ and any intermediate party $U_i, i \in [1,n-1]$ is stated in Procedure \ref{algo:rel11} and Procedure \ref{algo:rel12} respectively.
\begin{proc}[H]
    \SetKwInOut{Input}{Input}
    \SetKwInOut{Output}{Output}

    \caption{Establishing Payment Contract: Second Round of Locking Phase for $U_0$ }
        \label{algo:lock3}
        
\textbf{Input}: $(H,Y,\alpha_0,t_0)$  \\

\If{($U_1$ has agreed to form the contract) and (current\_time not close to contract expiration time)}
{
  $remain(U_{0},U_{1})=remain(U_0,U_{1})-\alpha_0$\\
  establish $Payment\_Contract(H,Y,t_0,\alpha_0)$ with $U_{1}$\\
 }
 \Else
 {

   abort
 }
 \end{proc}
 \begin{proc}[H]
    \SetKwInOut{Input}{Input}
    \SetKwInOut{Output}{Output}

    \caption{Establishing Payment Contract: Second Round of Locking Phase for $U_i, i \in [1,n-1]$ }
        \label{algo:lock4}
        
\textbf{Input}: $(H,Y,\alpha_i,t_i)$  \\

\If{$t_{i-1}\geq t_i +\Delta$ and $\alpha_i\stackrel{?}{=}\alpha_{i-1}+fee(U_i)$ and ($U_{i+1}$ has agreed to form the contract) and (current\_time not close to contract expiration time)}
{
  $remain(U_{i},U_{i+1})=remain(U_i,U_{i+1})-\alpha_i$\\
  establish $Payment\_Contract(H,Y,t_i,\alpha_i)$ with $U_{i+1}$\\
 }
 \Else
 {
 abort
 }
 \end{proc}

 \begin{proc}[!ht]
    \SetKwInOut{Input}{Input}
    \SetKwInOut{Output}{Output}

    \caption{Release\_Phase for $U_n$}
        \label{algo:rel11}
        
\textbf{Input}: Message $M$, time bound $\delta$  \\

\If{$M\stackrel{?}{=}Payment\_Contract(H,Y,\alpha',t')$ and $current\_clock\_time-t_{n}^{form}\leq \delta$}
{
Parse $M$ and retrieve $(H,Y,\alpha',t')$\\
   \If{ $t'\geq t_{now}+ \Delta$ and $\alpha'=\alpha$}
   {
      $z=x$\\
      
    }
    \Else
    {
      $z=r$\\
       
     }
   }
   \Else{
     $z=r$\\
   }
      Release $z$ to $U_{n-1}$\\
      \If{$current\_time < t_{n-1}$}
      {
         
         \If{$U_n$ and $U_{n-1}$ mutually agree to terminate \emph{Payment Contract and Cancellation Contract}}
         {
         
         \If{z=x}
         {
        $remain(U_n,U_{n-1})=remain(U_n,U_{n-1})+\alpha+\textrm{tgp}_{n-1}$\\         
        }
        \Else
        {
          $remain(U_n,U_{n-1})=remain(U_n,U_{n-1})+\textrm{tgp}_{n-1}$\\         
          $remain(U_{n-1},U_{n})=remain(U_{n-1},U_{n})+\alpha$\\         
        }
           
           }
           \Else
           {
  $U_n$ goes on-chain for settlement by releasing preimage $z$.           

         }
         
      }
      \Else
      {
      $U_{n-1}$ goes on-chain for settlement, claims $(\alpha+\textrm{tgp}_{n-1})$. \\
      $z=null$\\
      }
                    Call Release\_Phase($U_{n-1},z$)\\

   \end{proc}
  \begin{proc}
    \SetKwInOut{Input}{Input}
    \SetKwInOut{Output}{Output}

    \caption{Release\_Phase for $U_i, i \in [1,n-1]$}
        \label{algo:rel12}
        
\textbf{Input}: $z$  \\

      Release $z$ to $U_{i-1}$\\
      \If{$z\neq null$ and $current\_time < t_{i-1}$}
      {
         
         \If{$U_i$ and $U_{i-1}$ mutually agree to terminate \emph{Payment Contract and Cancellation Contract}}
         {
        \If{z=x}
         {
        $remain(U_{i},U_{i-1})=remain(U_{i},U_{i-1})+\alpha_{i-1}+\textrm{tgp}_{i-1}$\\         
        }
        \Else
        {
          $remain(U_i,U_{i-1})=remain(U_i,U_{i-1})+\textrm{tgp}_{i-1}$\\         
          $remain(U_{i-1},U_i)=remain(U_{i-1},U_i)+\alpha_{i-1}$\\         
        }

           }
           \Else
           {
  $U_i$ goes on-chain for settlement by releasing preimage $z$.\\

         }
         }
         
\Else
{
      $U_{i-1}$ goes on-chain for settlement after elapse of locktime $t_{i-1}$, claims $(\alpha_{i-1}+\textrm{tgp}_{i-1})$. \\
}

                    Call Release\_Phase($U_{i-1},z$)\\

   \end{proc}


%
\textbf{Safeguarding against Reverse-Griefing.} Any request for off-chain termination of contract by a party $U_i, i \in [1,n-1]$, without providing any of the preimage, will not be accepted by $U_{i-1}$ unless $U_i$ is compensating it for the loss of time. If the party $U_{i-1}$ mutually terminates the contract without the knowledge of any of the preimage before elapse of locktime, it is quite possible $U_{i-2}$ may refuse to cancel the contract and wait for the contract to expire. This might lead to the problem of \emph{reverse-griefing} where $U_{i-1}$ loses funds. Hence to safeguard itself, a party will agree to terminate the contract off-chain either on receiving griefing-penalty or on receiving any one of the preimage.


\section{Security Analysis}
\label{secgp1}


\begin{theorem}
\textbf{(Guaranteed compensation for an honest
node).} Given a payment request $(U_0,U_n,\alpha_0)$ to be transferred via path  $P=\langle U_0, U_1,\ldots,U_n \rangle$, if at least one party $U_{k}, k \in [1,n]$ mounts griefing attack then any honest party $U_j \in P, j \in [0,k-1]$ will earn compensation, without losing any funds in the process. 
\label{th11}
\end{theorem}

\textbf{Proof}:

We consider the worst-case in which we assume only a single node is honest in a path and rest of the nodes acts maliciously. We note that if fewer number of parties are corrupted, the honest nodes still interact with malicious neighbours and hence they get reduced to cases mentioned here. In particular, we analyze interactions between honest and dishonest parties in our system and ensure that honest parties do not get cheated and get their due.

\begin{itemize}[leftmargin=*]

\item \textit{\underline{Case 1 : $U_0$ is honest}} \\  In the \texttt{Pre-processing Phase}, the honest sender builds the onion packets containing terms of contract which are propagated through the nodes in the path. While the values in the packets are contingent to the values sent by R, honest S is in no position to verify it. At this point, S just follows the protocol. 


In the \texttt{Two Round Locking Phase}:
\begin{itemize}[leftmargin=*]
\item \emph{Establishing the Cancellation Contract}: Since $U_0$ is the last party to receive the contract, in case any party $U_i, i \in [1,n]$ griefs, then it will end up paying a cumulative penalty of $\gamma.(\Sigma_{j=1}^{i-1} (\alpha_j t_j)+(\alpha_0+\psi)t_0)$, whereby $U_0$ earns a compensation of $\gamma(\alpha_0+\psi)t_0$.

\item \emph{Establishing the Payment Contract}:  $U_0$ may not be able to forward the contract to $U_{1}$ if there is discrepancy in the terms of the outgoing contract or if $U_{1}$ has stopped responding. Since $U_n$ is dishonest, it will not release the preimage $r$ for cancelling the contracts established in the first round. As per the terms of the contract, after the elapse of locktime $t_{n-1}$, it pays a cumulative griefing-penalty $\gamma.(\Sigma_{j=1}^{n-1} (\alpha_j t_j)+(\alpha_0+\psi)t_0)$ to $U_{n-1}$. This money is used for compensating rest of the parties, starting from $U_{0}$ till $U_{n-1}$. Even if $U_1$ griefs, as per the terms of the contract it has to pay the required compensation $\gamma(\alpha_0+\psi)t_0$ to $U_0$. Hence $U_0$ will not lose funds. 
\end{itemize}

In the \texttt{Release Phase}, a similar argument can be given. Assuming that two round locking phase got executed successfully, if $U_{1}$ is not able to release the preimage corresponding to either cancellation hash or payment hash before the elapse of contract's locktime, it will have to pay a penalty to $U_0$.


\item \textit{\underline{Case 2 : An intermediate node $U_i, i \in [1,n-1]$ is honest}} \\ For the \texttt{Pre-processing Phase}, if $U_i$ does not receive the onion packet, the payment won't get instantiated.

In the \texttt{Two Round Locking Phase}:
\begin{itemize}[leftmargin=*]
\item \emph{Establishing the Cancellation Contract}: $U_i$ may not be able to forward the contract to $U_{i-1}$ if the latter stops responding. Since $U_n$ is dishonest, it will not release the preimage $r$ for cancelling the contracts established in the first round. As per the terms of the contract, after the elapse of locktime $t_{n-1}$, it pays a cumulative griefing-penalty $\gamma.(\Sigma_{j=1}^{n-1} (\alpha_j t_j)+(\alpha_0+\psi)t_0)$ to $U_{n-1}$. Even if $U_{i+1}$ griefs, as per the terms of the contract it has to pay the required compensation $\gamma( (\alpha_0+\psi)t_0+ \Sigma_{j=1}^{i} (\alpha_j t_j))$ to $U_i$. Since no contract has been established between $U_i$ and $U_{i-1}$, $U_i$ retains the entire compensation.

\item \emph{Establishing the Payment Contract}: $U_i$ may not be able to forward the contract to $U_{i+1}$ if  $U_{i+1}$ stops responding. The same logic stated for \emph{cancellation contract} holds true except now $U_i$ can retain $\gamma \alpha_i t_i$ as compensation and forward the rest of the amount to node $U_{i-1}$. Even if $U_{i-1}$ doesn't respond and wait for the locktime $t_{i-1}$ to elapse, $U_i$ will not lose funds.
\end{itemize}

In the \texttt{Release Phase}, $U_i$ can be griefed in the following ways:
\begin{itemize}[leftmargin=*]
\item $U_{i+1}$ withholds the preimage (either cancellation or payment) from $U_{i}$ and waits for the contract locktime to expire. In that case, $U_{i+1}$ has to pay compensation of $\gamma ((\alpha_0+\psi)t_0+ \Sigma_{j=1}^{i} (\alpha_j t_j))$ to $U_{i}$. Even if $U_{i-1}$ reverse-griefs, $U_i$ will be able to compensate without incurring any loss. 
\end{itemize}
\item  \textit{\underline{Case 3 : $U_n$ is honest}} \\ Receiver $U_n$ initiates the release of preimage. It will resolve the payment within a bounded amount of time either by releasing the preimage for payment hash or cancellation hash, as per the situation. $U_{n-1}$ cannot reverse-grief and force receiver to pay a griefing-penalty. 

\end{itemize}

\begin{theorem}
\label{priv}
\textbf{(Payer and Payee's Privacy).} Given the information of griefing-penalty in the off-chain contract, an intermediate node cannot infer its exact position in the path for routing payment.  
\label{21}
\end{theorem}
\textbf{Proof}: For routing payment of amount $\alpha$ from $U_0$ to $U_n$ via intermediaries $U_i, i \in [1,n-1]$, several instances of off-chain contract is established across the payment channels. The amount locked by party $U_j$ and $U_{j+1}$ in their off-chain contract is $\alpha_j$ and $\gamma((\psi + \alpha_0)t_0+\Sigma_{k=1}^{k=j} \alpha_j t_{j}), j \in [0,n-1]$, respectively. Let us assume that there exists an algorithm $\tau$ which reveals the exact position of any intermediate node $D : D \in \{U_1,U_2,\ldots,U_{n-1}\}$ in the path. This implies that given the information of cumulative griefing-penalty mentioned in the contract, it can distinguish between the penalty charged by channel $(U_j,U_{j+1}), j \in [1,n-1]$ and penalty charged by channel $(U_0,U_1)$, which is $\gamma ((\psi+\alpha_0)t_0)$. However, the routing attempt cost $\psi$, was added by node $U_0$ as an extra compensation charged to cover up for routing attempt expense as well as hiding its identity from its next neighbour. This is information is private and not known by any node except $U_0$. Additionally, the value of $\psi$ is set such that $\psi t_{0} \geq  \alpha ((k+1) t_0+\Sigma_{l=1}^{k} l\Delta), k \in \mathbb{N}$. Any number being selected from $\mathbb{N}$ being equiprobable, the probability of distinguishing becomes negligible. 

Note: \emph{In practical application, there is a limit on the routing attempt fee which a sender can charge. The set from which $k$ is selected is significantly smaller compared to $\mathbb{N}$. But even under such circumstances, the best inference made by any intermediate node $U_j$ about its location is that it is located at position $(j+k)$ where $\psi t_0 \geq \alpha. t_0 + \alpha.(t_0+\Delta) + \alpha.(t_0+2\Delta) + \ldots + \alpha.(t_0+k\Delta), k$ acts as the blinding factor.}

\section{Performance Evaluation}
\label{performance}
\subsection{Analysis of Profit earned by eliminating a Competitor from the Network}
\label{profit}

The motivation of the griefer is to eliminate a competitor. It will try to exhaust all the channel capacity of the victim and force all transactions to be routed through itself, with the expectation to earn processing fee.

\emph{Return on Investment} or \emph{RoI} is the profit earned by the attacker with respect to the investment made in the network. Here investment means the liquidity utilized by the attacker for simulating an attack. In Lightning Network, the \emph{RoI} of a node processing transaction request is calculated as follows:
\begin{equation}
\label{roi}
\begin{matrix}
profit\_processed =N_{tx}(base\_fee+fee\_rate* tx\_value)\\
RoI=profit\_processed-total\_griefing\_penalty
\end{matrix}
\end{equation}
\emph{profit\_processed} is calculated based on \cite{fee1}, \cite{fee2}. $N_{tx}$ is the total number of transactions processed by the node and $tx\_value$ is the amount transferred from payer to payee. \textit{total\_griefing\_penalty} is the penalty required to pay as compensation to the affected parties upon mounting griefing-attack. For \emph{HTLC}, the \textit{total\_griefing\_penalty} is 0 since there is no concept of penalizing the attacker. Hence the node always earns a non-negative \emph{RoI}.

For \emph{HTLC-GP}, if the node mounts a griefing attack, it has to pay a griefing-penalty proportional to the collateral locked for the given timeperiod. If the $\textit{total\_griefing\_penalty}$ exceeds the \textit{profit\_processed}, the node incurs a loss.  In the next section, we define two strategies which can be opted by the attacker. Based on these strategies, we compare the \emph{Return on Investment} obtained for \emph{HTLC} and \emph{HTLC-GP} for a given budget.

\begin{figure}[!ht]
    \centering
    \subfloat[Attacker establishes two edges with a targeted source and targeted sink connected to the victim]{{\includegraphics[width=6cm]{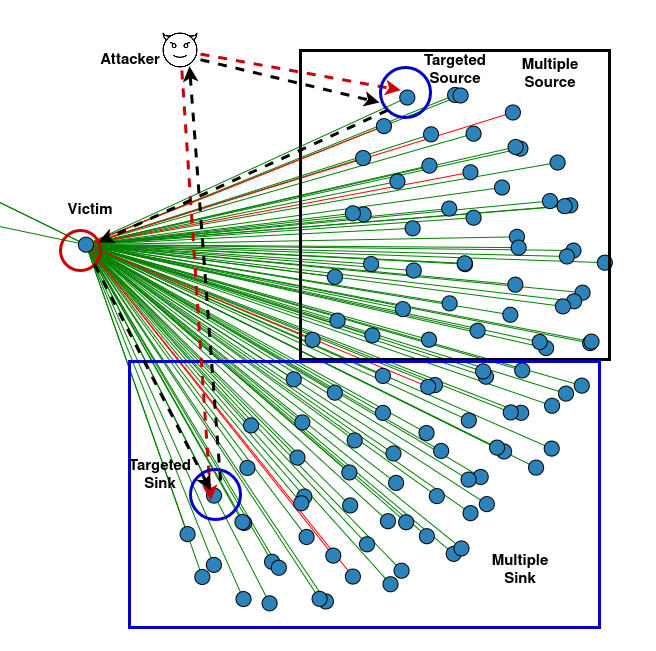} }}%
    \qquad
    \subfloat[$Attacker$ uses existing channel for mounting the attack]{{\includegraphics[width=6cm]{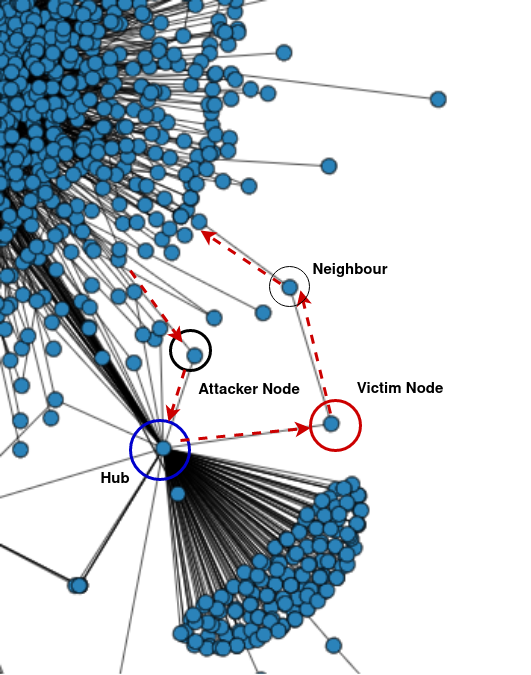} }}%
   \caption{Snapshot of the network on $19^{th} May, 2020$}
\label{Star21}
\end{figure}

\subsubsection{Attacking Strategies}
\label{a1}
\subsubsection*{(a) Attacker establishes additional channels}

Nodes with \emph{high betweenness centrality} tend to act as intermediaries for routing payments. The attacker selects such nodes as its victim. We illustrate the situation by studying the structure of an instance of Lightning Network. The snapshot taken on $19^{th}$ May, 2020, in Fig. \ref{Star21}(a). Nodes marked as \texttt{Targeted Source} and \texttt{Targeted Sink} routes their payment via node \texttt{Victim}. A malicious node establishes new channels with the \texttt{Targeted Source} and \texttt{Targeted Sink}. It selects the route \texttt{Attacker}$\rightarrow$\texttt{Targeted Source}$\rightarrow$\texttt{Victim}$\rightarrow$\texttt{Targeted Sink}$\rightarrow$\texttt{Attacker}, sends self-payment requests and mounts griefing attack. The path \texttt{Targeted Source}$\rightarrow$\texttt{Victim}$\rightarrow$\texttt{Targeted Sink} gets blocked. All the payments from \texttt{Targeted Source} gets routed through the path \texttt{Targeted Source}$\rightarrow$\texttt{Attacker}$\rightarrow$\texttt{Targeted Victim}.

\subsubsection*{(b) Attacker uses existing channels}
\label{a2}

In the previous strategy, the attacker had to establish channels before mounting the attack. To avoid the cost of establishing new channels, the attacker makes use of its existing payment channels to block payments received by its competitor. Illustrating the attack on the same instance of Lightning Network, as shown in Fig. \ref{Star21}(b), we consider that the node marked as \texttt{Hub} routes all the payment request via \texttt{Victim node} and ignores sending any payment via \texttt{Attacker node}. In order to steal payments being routed via \texttt{Victim node}, it selects the route \texttt{Attacker Node}$\rightarrow$\texttt{Hub}$\rightarrow$\texttt{Victim Node}$\rightarrow$\texttt{Neighbour}\ldots$\rightarrow$\texttt{Attacker Node} for self-payment and mounts griefing attack. The path \texttt{Hub}$\rightarrow$\texttt{Victim Node}$\rightarrow$\texttt{Neighbour}$\ldots\rightarrow$\texttt{Attacker Node} gets blocked. Now \texttt{Hub} will be forced to route such payment request through \texttt{Attacker Node}. 

%

%
\subsubsection{Experimental Analysis}
\label{evaluate}

\textbf{Setup}: For our experiments, we use Python 3.8.2 and NetworkX, version 2.4 \cite{hagberg2008exploring} - a Python package for analyzing complex networks. System configuration
used is Intel Core i5-8250U CPU, Operating System: Kubuntu 20.04, Memory: 7.7 GiB of RAM. The code for our implementation is available on GitHub \footnote{\href{https://github.com/subhramazumdar/GriefingPenaltyCode}{https://github.com/subhramazumdar/GriefingPenaltyCode}}. From
the dataset mentioned in \cite{mizrahi2020congestion}, we took twelve snapshots of Bitcoin Lightning Network over a year, starting from \emph{September, 2019}. Each snapshot provides information regarding the  public key of the nodes and the aliases used. The topology of the network is represented in the form of channels, represented as pair of public keys along with the channel capacity and the channel identifier. Each node of the channel follows a node policy which mentions about the base fee in millisatoshi, fee rate per million (in millisatoshi) and time\_lock\_delta. The capacity denotes the money deposited in the channel and not the balance of individual parties involved in opening of the channel. Thus each snapshot of Lightning Network undergo preprocessing where we filter out channels which is marked as disabled. Next, we select the largest connected component in the network. Since our proposed strategy for countering griefing attack requires both the parties to fund the channel, we divide the capacity of the channel into equal halves and allocate each half as the balance of a counterparty. The preprocessed graphs are used for evaluating both \emph{HTLC} and \emph{HTLC-GP}. 

\textbf{Designing transaction set}: The best way to approximate the maximum value routed through a node is to map it into a flow problem and compute the maximum flow \cite{ford2009maximal} from multiple payers/sources to multiple payees/sinks. The amount of flow across each channel is the upper bound on the number of transactions being processed. If the attacker manages to block at least one path connecting a payer and payee, then the payer will route its transaction via the attacker. Since it is easier to analyze the situation in a \emph{hub-and-spoke} network, we select a subgraph of LN having a similar structure. Nodes with high \emph{betweenness centrality} \cite{rohrer2019discharged} have high probability of being a potential \emph{hub node}. We select such a hub node where a subset of the pendant nodes connected to the hub forms the set of source and rest of the neighbours form the sink. Once a maximum flow is computed, the flow through a channel connecting a source to hub and through the channel from hub to a sink forms the maximum valued payment that can be routed through the path. The attacker targets such a source-sink pair with the hub acting as its victim. The attacker selects the victim, blocks its channels. Once the maximum flow across such source-sink pairs gets computed, the attacker checks the fraction of flow which gets routed through itself. To get an estimate of transaction set size, it divides the flow by the amount per transaction. \emph{Return on Investment} can be calculated for all such transactions based on Eq. \ref{roi}.  


\textbf{Data Used}: We vary the range of transaction amount between \emph{1 satoshi} to \emph{100000 satoshi} \cite{beres2019cryptoeconomic}, increasing the amount by multiple of 10. For the attack involving establishment of new channels by the attacker, we vary the level of budget of the adversary as \emph{3000 satoshi, 30000 satoshi, 300000 satoshi, \ldots, 3 BTC}. For the attack involving use of existing channels by the attacker, we vary the level of budget of the adversary as \emph{3000 satoshi, 30000 satoshi, 300000 satoshi, \ldots, 0.03 BTC}. Increasing budget beyond 0.03 BTC is of no use since substantial amount of budget remains unutilized after this point. Note that here budget allotted is utilized by the attacker for instantiating payment and locking cumulative griefing-penalty, ignoring the cost of establishing the channels in the network.  

\subsubsection*{Simulation Result}
 The \emph{Return on Investment (RoI)} is measured in log-scale. For negative \emph{RoI}, we use log-modulus transformation \cite{john1980alternative}.

\begin{figure}[!ht]
    \centering
    \subfloat[RoI vs Average value per transaction: Fixed Budget - 0.03 BTC ]{{\includegraphics[width=7cm]{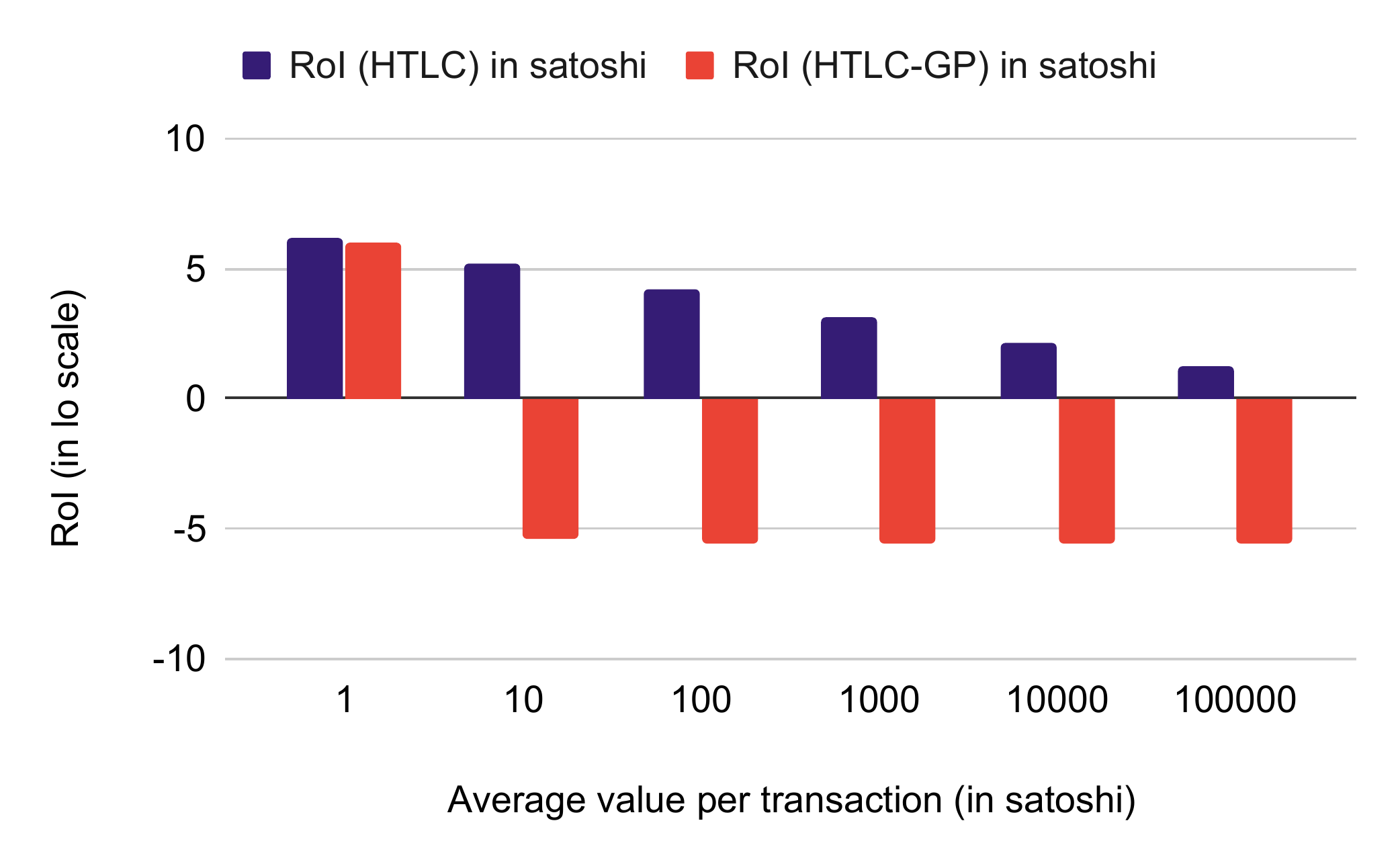} }}%
    \qquad
    \subfloat[RoI vs Budget: Fixed Average value per transaction - 10000 satoshi]{{\includegraphics[width=7cm]{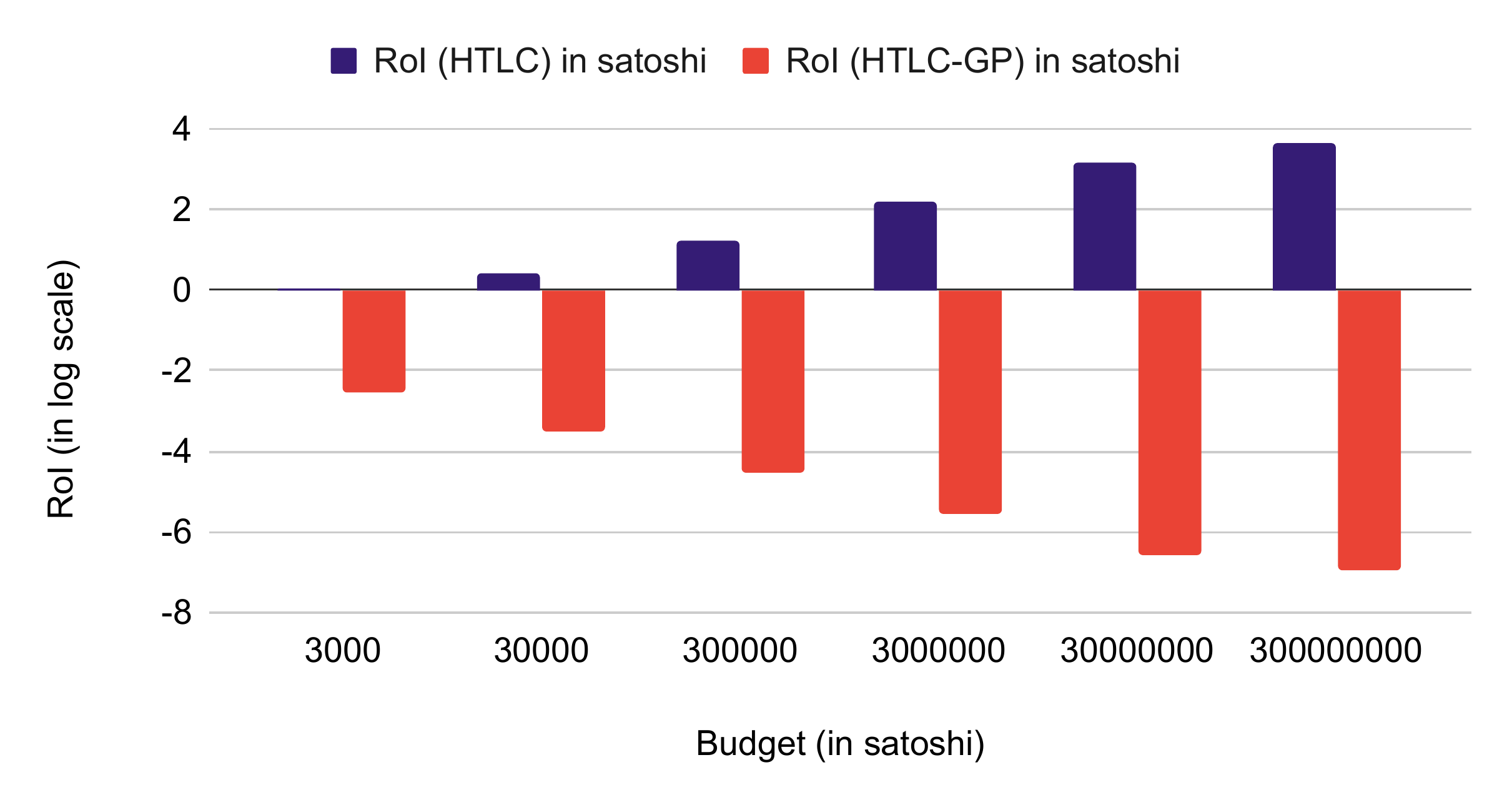} }}%
        \qquad
    \subfloat[RoI vs Rate of Griefing-Penalty: Fixed Average value per transaction - 10000 satoshi]{{\includegraphics[width=7cm]{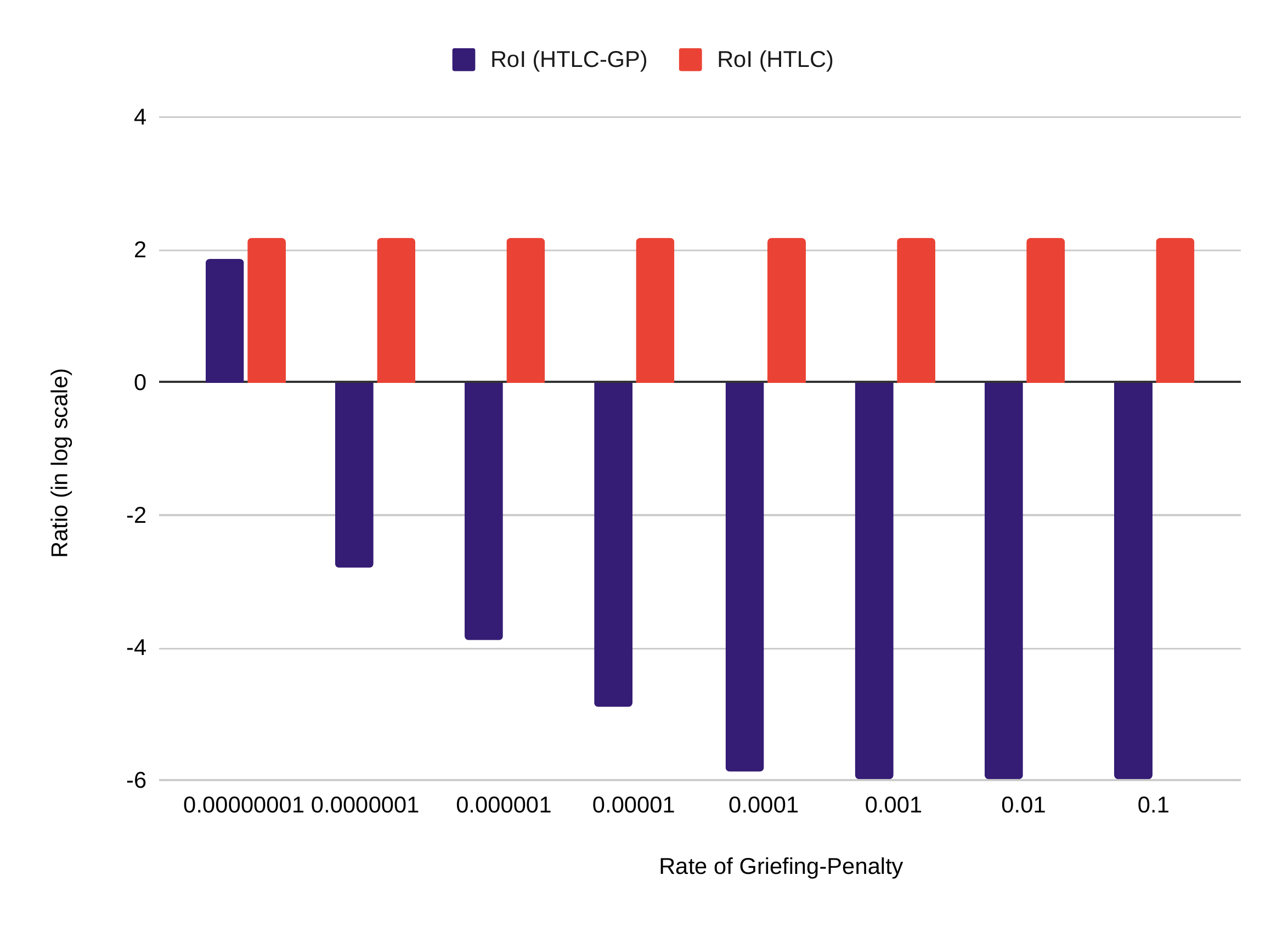} }}%

    \caption{When Attacker uses new channels for mounting the attack }%
    \label{fig:example12}%
\end{figure}

\begin{itemize}[leftmargin=*]

\item \emph{Using attacking strategy 1}: The first result \emph{RoI vs Average value per transaction}, for a fixed budget of 0.03 BTC and fixed rate of griefing-penalty of 0.001 per minute, shows that as the average value of each transaction increases, the processing fee earned by the attacker decreases due to decrease in the maximum number of payments processed for \emph{HTLC}. However, for \emph{HTLC-GP}, as the average value per transaction increases, \emph{RoI} becomes negative as the processing fee earned becomes negligible compared to penalty incurred, as shown in Fig. \ref{fig:example12}(a). 
 
The second result \emph{RoI vs Budget}, for a fixed average value of transaction of 10000 satoshi and fixed rate of griefing-penalty of 0.001 per minute, shows that as the budget of the attacker increases, the processing fee earned by the attacker increases linearly. But a reverse trend is observed for \emph{HTLC-GP}. \emph{RoI} decreases linearly, as shown in Fig. \ref{fig:example12}(b). This is because is the amount of cumulative penalty is directly related to total collateral locked by the attacker. 

The third result \emph{RoI vs Rate of Griefing-Penalty}, for a fixed average value of transaction of 10000 satoshi and fixed budget of 0.03 BTC, the return on investment for \emph{HTLC} remains constant since rate of griefing-penalty has no impact in this case. But the loss incurred increases with increase in $\gamma$, as observed for \emph{HTLC-GP} in Fig. \ref{fig:example12}(c). 

\begin{figure}[!ht]
    \centering
    \subfloat[RoI vs Average value per transaction: Fixed Budget - 0.03 BTC  ]{{\includegraphics[width=7cm]{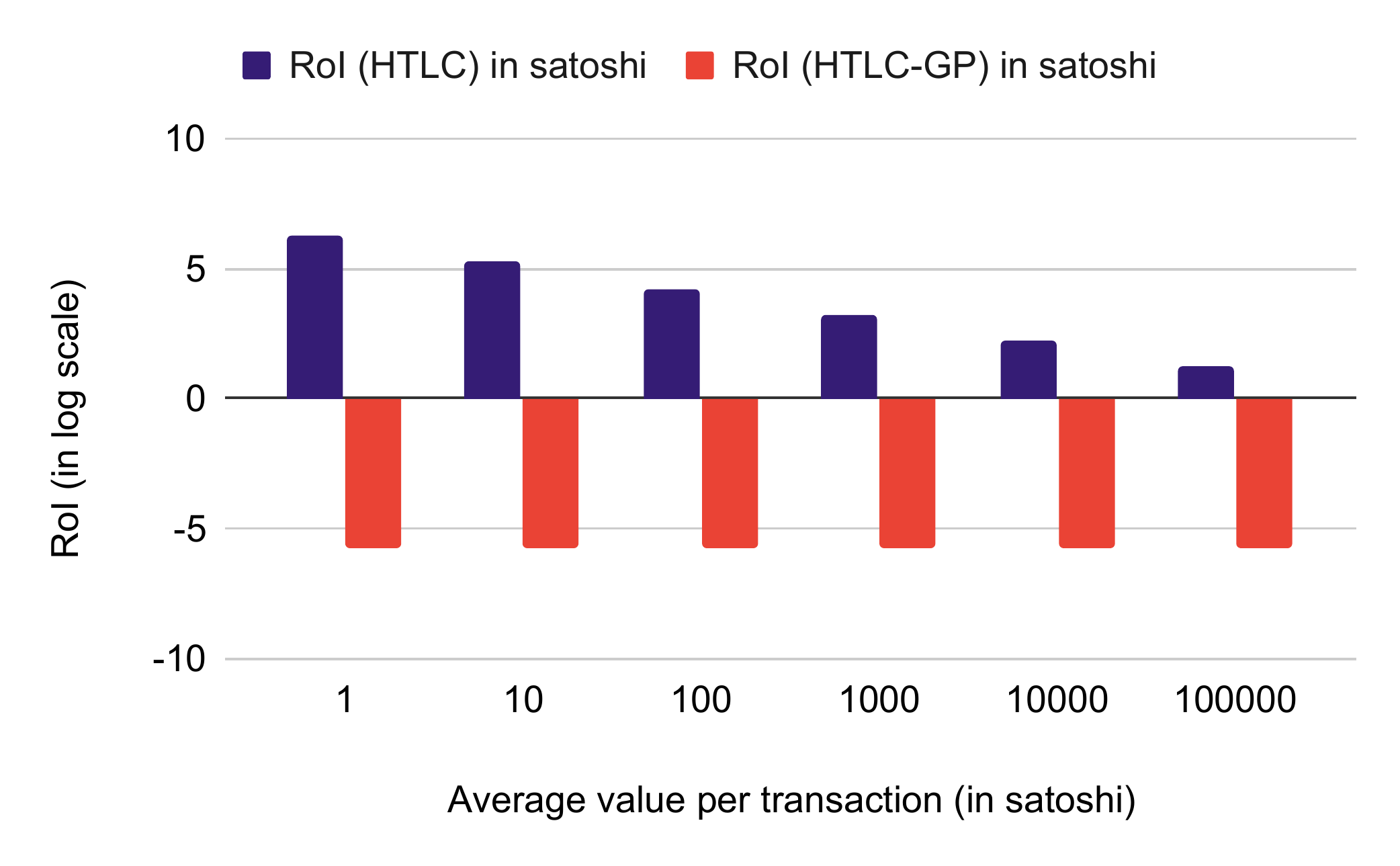} }}%
    \qquad
    \subfloat[RoI vs Budget: Fixed Average value per transaction - 10000 satoshi]{{\includegraphics[width=7cm]{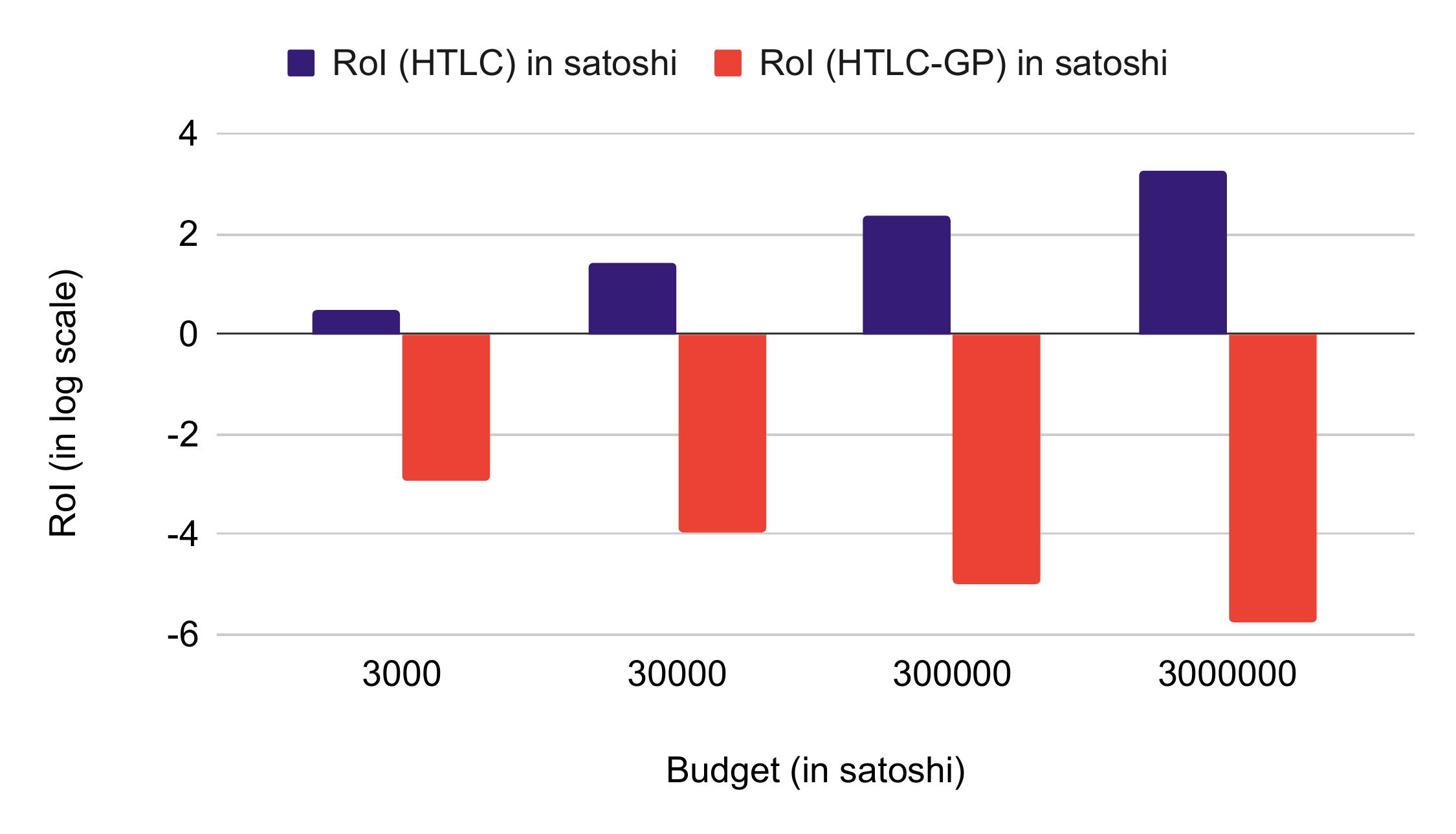} }}%
            \qquad
    \subfloat[RoI vs Rate of Griefing-Penalty: Fixed Average value per transaction - 10000 satoshi]{{\includegraphics[width=7cm]{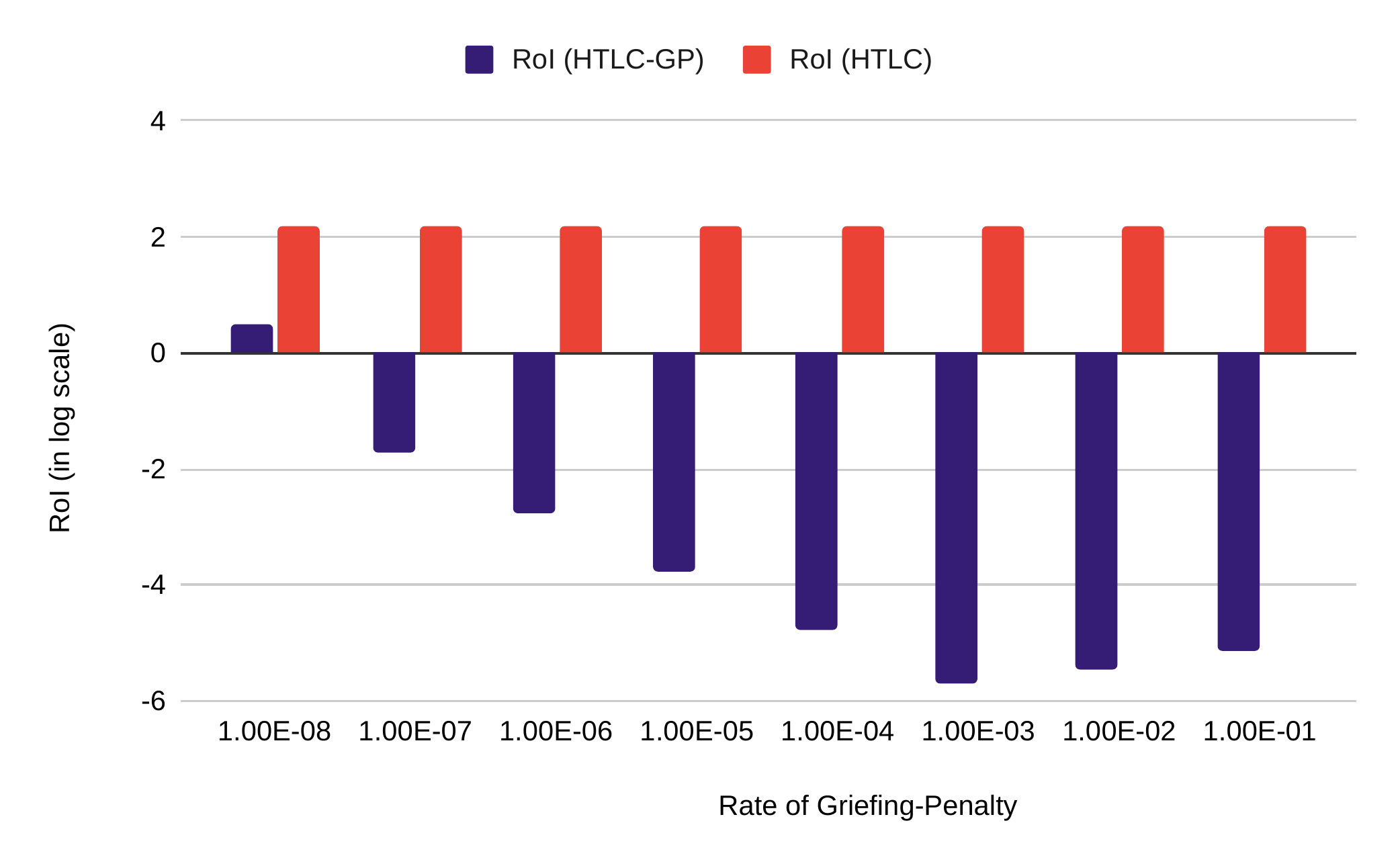} }}%

    \caption{When Attacker uses existing channels for mounting the attack }%
    \label{fig:example13}%
\end{figure}

\item \emph{Using attacking strategy 2}: For a fixed budget of the attacker, loss incurred for \emph{HTLC-GP} using second attacking strategy is higher than the first attacking strategy for all the three cases, as shown in Fig Fig. \ref{fig:example13}(a), Fig. \ref{fig:example13}(b) and Fig. \ref{fig:example13}(c). This is because average path length for self-payment being around 6.5 compared to the first attacking strategy, where the average path length remains fixed at 4.



\end{itemize}

\subsection{Investment made by attacker for stalling the network}
\label{vary1}
For a path of length $n$, the cumulative griefing-penalty is $\gamma((\psi+\alpha_0)t_0 + \Sigma_{j=1}^{n-1}\alpha_j t_j)$ for transferring an amount of $\alpha_{n-1}$ from sender $U_0$ to receiver $U_n$. In case of \emph{HTLC}, for blocking liquidity of at least $\alpha_{n-1}$ in each of the $n$ channels, the attacker needs to invest $\alpha_0$ and execute a self-payment. In case of \emph{HTLC-GP}, in order to execute a self-payment of $\alpha_0$, the attacker needs to invest $\alpha_0+ \gamma((\psi+\alpha_0)t_0 + \Sigma_{j=1}^{n-1} \alpha_j t_j)$. If we take the ratio of the investment made for \emph{HTLC} and investment made by attacker for \emph{HTLC-GP} for a fixed transaction value,
\begin{equation}
\begin{matrix}
\frac{\alpha_0}{\alpha_0+ \gamma((\psi+\alpha_0)t_0 + \Sigma_{j=1}^{n-1}\alpha_j t_j)}
\leq  \frac{1}{1+ \gamma(t_0 + \Sigma_{j=1}^{n-1}\frac{\alpha_j}{\alpha_0} t_j)}\\

\end{matrix}
\end{equation}
\begin{itemize}[leftmargin=*]
\item \emph{Path Length}:  Keeping $\gamma$ and $\alpha_n$ fixed, with increase in path length $n$, the ratio will be strictly less than 1 for any $n>1$, since $\gamma(t_0 + \Sigma_{j=1}^{n-1}\frac{\alpha_j}{\alpha_0} t_j)>0$.    
\item \emph{Rate of Griefing-Penalty}: Keeping  path length $n$ and $\alpha_n$ fixed, with increase in rate of griefing-penalty, the ratio will be strictly less than 1 for any value of $\gamma \in (0,1)$ since $\gamma(t_0 + \Sigma_{j=1}^{n-1}\frac{\alpha_j}{\alpha_0} t_j)>0$.   
\end{itemize}

%
%
%
%
%

\textbf{Evaluation.} We use the same experimental setup and graph instances as in Section \ref{evaluate}.

\begin{figure}[!ht]
    \centering
    \subfloat[Impact of path length on the investment made by attacker for launching griefing attack (HTLC-GP vs HTLC) ]{{\includegraphics[width=9cm]{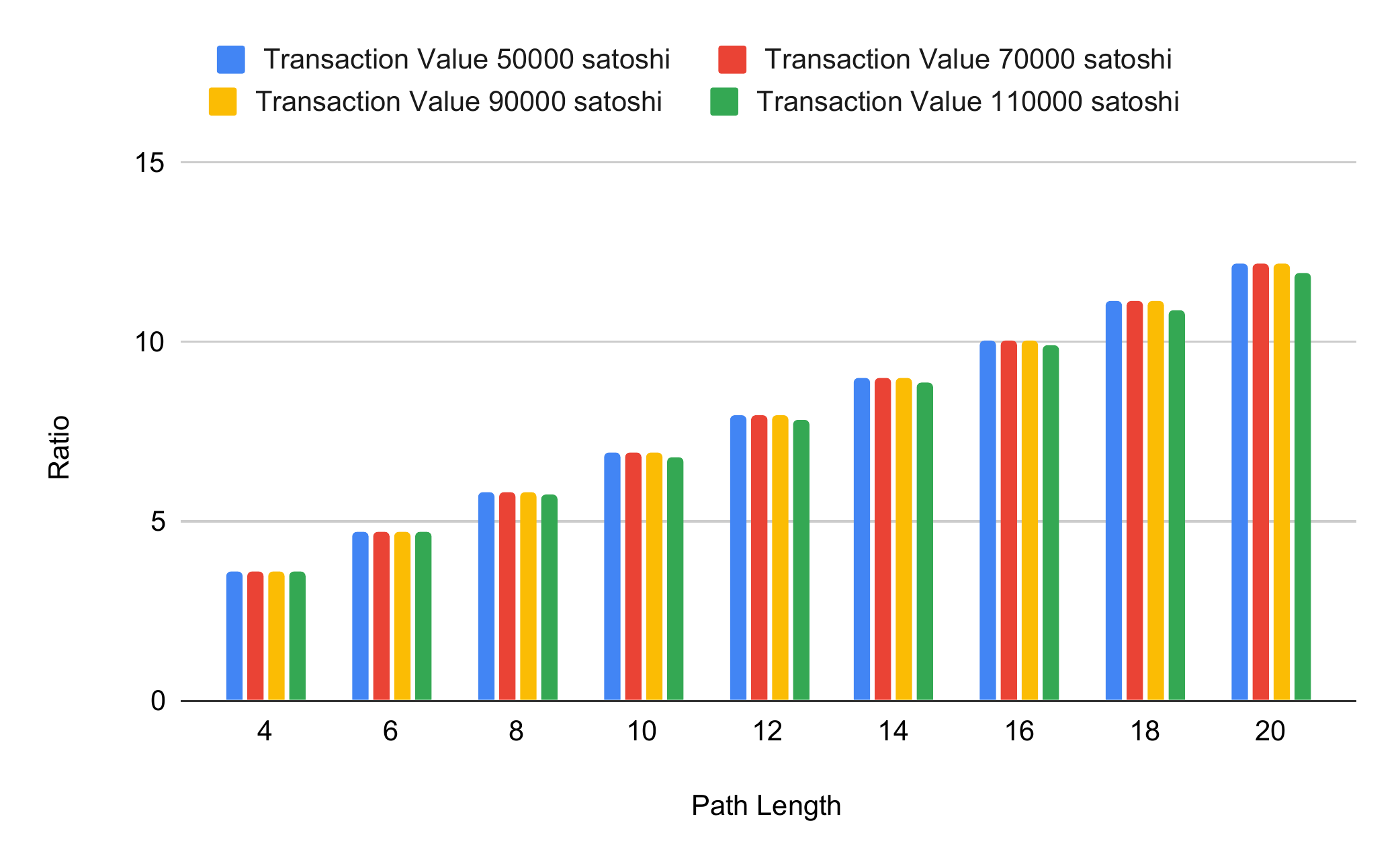} }}%
    \qquad
    \subfloat[Impact of rate of griefing-penalty on the investment made by attacker for launching griefing attack (HTLC-GP vs HTLC)]{{\includegraphics[width=9cm]{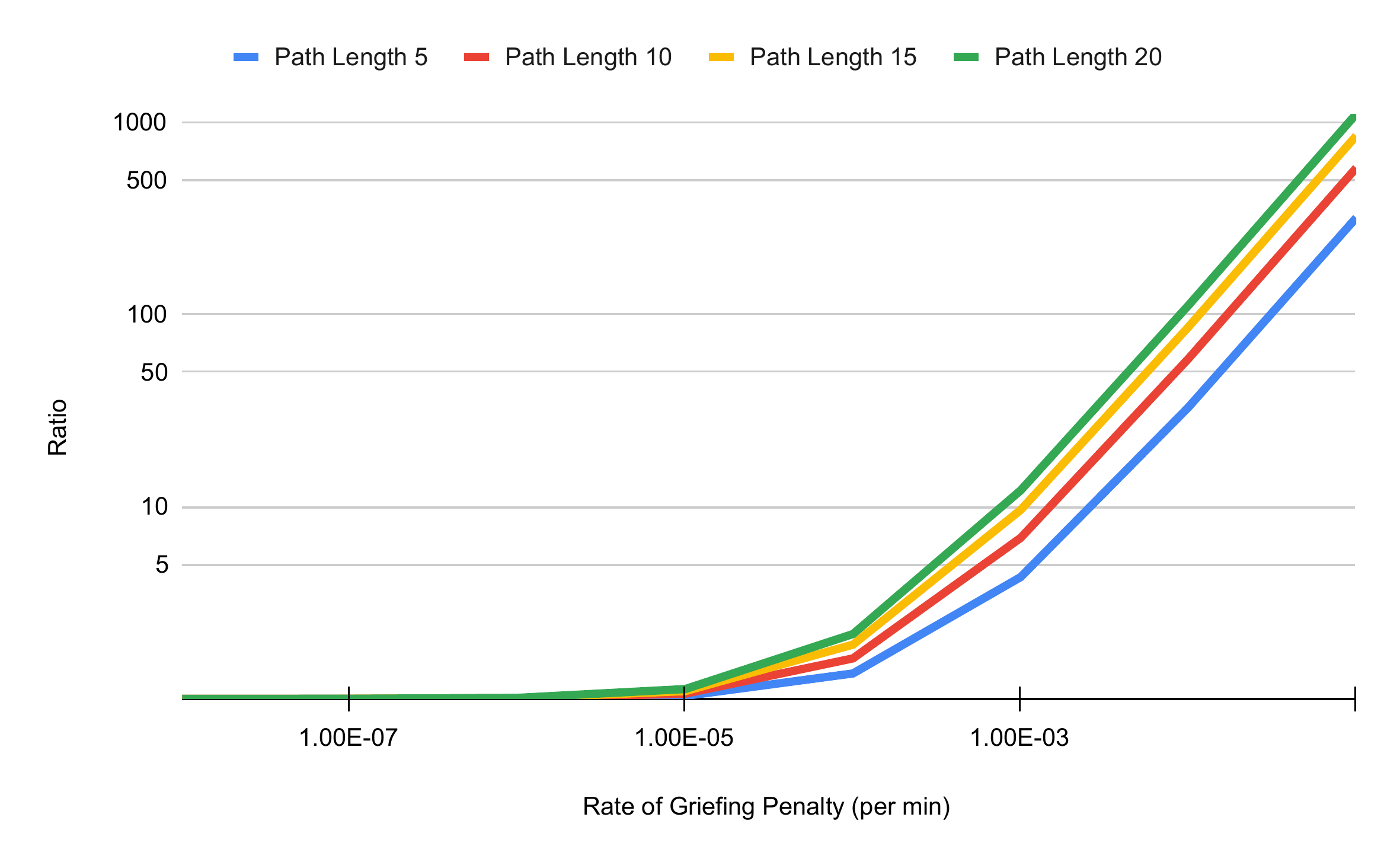} }}%

    \caption{ Investment made by attacker (HTLC vs HTLC-GP)}%
    \label{Pathl}%
\end{figure}

\begin{itemize}

\item \emph{Impact of Path Length.} For a given transaction value and fixed rate of griefing-penalty set to 0.001 per minute, we vary the path length in the range from 4 to 20, and transaction value as \emph{50000 satoshi, 70000 satoshi, 90000 satoshi, 110000 satoshi}. The ratio of the adversary budget needed for mounting griefing attack in \emph{HTLC-GP} and the adversary budget needed for mounting griefing attack in \emph{HTLC} is around 4.7, when path length is 4 and around 12 when path length is 20. The ratio increases linearly with increase in path length, as observed in Fig. \ref{Pathl}(a). Upon varying the transaction value, we do not observe any change in this trend.

\item \emph{Impact of Rate of Griefing-Penalty.} For a fixed transaction value of 50000 satoshi and given path length, we vary the rate of griefing-penalty $\gamma$ in the range $\{10^{-8},10^{-7},10^{-6},\ldots,0.01,0.1\}$ and the path length as \emph{5,10,15,20}. The ratio of the adversary budget needed for mounting griefing attack in \emph{HTLC-GP} and \emph{HTLC} increases exponentially with increase in rate of griefing-penalty. The rate of increase in ratio is almost equal till $\gamma$ is $10^{-4}$, invariant of change in path length. When $\gamma>10^{-3}$, the rate of increase in the ratio is the lowest for path length of 5 and increases faster for path length of 20, as shown in Fig. \ref{Pathl}(b). 

The result shows that the investment made by the attacker for \emph{HTLC-GP} is higher than the investment made by the attacker for \emph{HTLC} thereby strongly disincentivizing griefing attack.

\end{itemize}

\section{Related Works}
\label{related}
 \begin{table*}[h]
\centering
  

 \scalebox{0.9}{
  \begin{tabular}{|p{2.5cm}|p{3.5cm}|p{2cm}|p{4.5cm}|p{2cm}|} 
    \hline
  &Countermeasure suggested &Privacy of payer/payee &Problem &Compensation for affected parties  \\
 \hline
Rohrer et al. \cite{rohrer2019discharged} &Limit the number of incoming channel &Yes &Adversary can split funds over multiple channels and mount the attack &None\\
Mizrahi et al. \cite{mizrahi2020congestion}&Faster resolution of HTLC &Yes &Synchronization problem, parties can cheat &None\\
Miller et al. ~\cite{miller2019sprites}&Constant locktime for payment&Violated  &Applicable for ethereum styled payment network &None\\
Egger et al. ~\cite{egger2019atomic}&Constant locktime for payment &Violated &Relationship anonymity in a path routing payment doesn't exist &None\\
Up-front payments ~\cite{russell}&Sender pays each party excess fee as compensation in case there is an attack &Violated  &High economic barrier for sender of payment &Yes\\
Reverse bonds~\cite{joost1} &Receiver pays penalty to each of the affected parties &Violated &Not defined properly, privacy of payment violated &Yes\\
Proof-of-Closure ~\cite{zmn} &Payment channels have soft timeout period and intiate closure in case of delay &Yes &Not effective, attacker can still jam the network &None\\
\hline

\end{tabular}
}
\vspace{0.2cm}
\caption{Summary of existing countermeasures}
\label{tab:rel}
\vspace{-0.2cm}
\end{table*}

Several ideas have been proposed for countering griefing attack. A limit on the number of incoming channel as well as the channel capacity was proposed in \cite{rohrer2019discharged} as countermeasure for node isolation attack. However, the attacker may split the funds over multiple identities and channels to bypass the restrictions imposed. Game theoretic approach for analyzing the strategies of attacker and defender was proposed in \cite{tochner2019hijacking}. Faster resolution of \emph{HTLC} has been stated in \cite{mizrahi2020congestion} as another method to avoid the disadvantage of having staggered locktime across payment channels. However, such a feature would violate the purpose of having \emph{HTLC} timeout which acts as a safety net against other possible malicious activities. All these payment protocols had a staggered locktime over each channel responsible for routing the payment. The collateral cost incurred for staggered locktime protocols is substantial. Sprites \cite{miller2019sprites}, an Ethereum styled payment network, first proposed the idea of using constant locktime for resolving payment. However, privacy was violated as the path information, identity of sender and receiver was known by all participants involved in routing the payment. A similar concept of reducing collateral cost using constant locktime contracts was proposed for Bitcoin-compatible payment networks in \cite{egger2019atomic}. However, it violated relationship anonymity and the proposed protocol is yet to be realized practically. 

Alternate mitigation strategies by incentivizing or punishing nodes have been stated in the past. Use of up-front payment was first proposed in \cite{russell}. In up-front payment, a party has to pay fee to the other party for accepting the \emph{HTLC}. An excess fee paid is returned back to the sender upon successful resolution of payment. This introduces a lot of economic barrier where up-front payment may exceed the transaction fee. For small valued payment, a large up-front payment is a serious problem. Later, in \cite{joost1}, the concept of reverse-bond was proposed which is similar to our proposed strategy. The counterparty accepting the \emph{HTLC} will have to pay a hold-fee on a per unit interval basis, as if it has rented the \emph{HTLC}. However, it has not been stated formally how this can be realized plus there is no way to track per unit interval in a decentralized asynchronous setup. Up-front payments has also been used for disincentivizing griefing attack in atomic swaps \cite{heilman2020arwen}. In \cite{zmn}, a proposal of Proof-of-Closure of channels was proposed, where by each \emph{HTLC} will have a hard timeout and a soft timeout period. However, a malicious node can setup several sybil nodes just for this purpose so that channel closure doesn't affect its normal activity in the network. Table \ref{tab:rel} provides a summary of the existing countermeasures and their disadvantages. Our proposed protocol addresses these shortcoming. 
%


\section{Conclusion \& Future Work}
\label{conclusion}
In this paper, we have proposed a strategy for mitigating griefing attack in Lightning Network by imposing penalty on the adversary. This increases the total cost for launching such an attack as well as compensates other nodes in the network affected by griefing. We have shown how our proposed strategy works in a timelocked payments by proposing a new  protocol \emph{HTLC-GP}. The proposed construction not only preserves privacy but also ensures that none of the honest intermediary present in the path gets affected due to imposition of penalty. 

As part of our future work, we would like to extend the concept of griefing-penalty to \emph{Atomic Cross Chain Swap}. A game-theoretic analysis for cross chain swaps using HTLC in \cite{xu2020game} states the locking collateral by both the parties results in higher success rate of transaction. However this protocol assumes both the parties lock same amount of collateral in a single smart contract belonging to either of the blockchain. We would like to study the impact of exchange rate volatility, locktime of contract on the cumulative griefing-penalty, with each party locking collateral in different contracts belonging to different blockchains.
\section*{Acknowledgement}
The authors are immensely grateful to Lightning Network Developers for their invaluable feedback.
\bibliography{PCN}

\end{document}